\documentclass{aa}  
\usepackage{amsmath}
\usepackage{graphicx}
\usepackage{pifont}
\usepackage[colorlinks=true,citecolor=blue]{hyperref}
\usepackage[export]{adjustbox}
\usepackage{txfonts}
\usepackage{xcolor}
\usepackage{soul}

\begin{document} 

   \title{The AIDA-TNG project: dark matter profiles and concentrations in alternative dark matter models}
    \titlerunning{Dark matter profiles in WDM and SIDM}
    \authorrunning{Despali et al.}

   \author{Giulia Despali
          \inst{1,2,3}\fnmsep\thanks{giulia.despali@unibo.it}, Carlo Giocoli\inst{2,3}, Lauro Moscardini\inst{1,2,3}, Annalisa Pillepich\inst{4}, Mark Vogelsberger\inst{5}, Massimo Meneghetti\inst{2}}

   \institute{Dipartimento di Fisica e Astronomia "Augusto Righi", Alma Mater Studiorum Università di Bologna, via Gobetti 93/2, I-40129 Bologna, Italy \label{1}
   \and INAF-Osservatorio di Astrofisica e Scienza dello Spazio di Bologna, Via Piero Gobetti 93/3, I-40129 Bologna, Italy \label{2}
   \and INFN-Sezione di Bologna, Viale Berti Pichat 6/2, I-40127 Bologna, Italy \label{3}
   \and Max-Planck-Institut für Astronomie, Königstuhl 17, 69117 Heidelberg, Germany \label{4}
      \and Department of Physics, Kavli Institute for Astrophysics and Space Research, Massachusetts Institute of Technology, Cambridge, MA 02139, USA \label{5}
        }

   \date{}

  \abstract {In the standard Cold Dark Matter (CDM) scenario, the density profiles of dark matter haloes are well described by analytical models linking their concentration to halo mass. Alternative scenarios, such as warm dark matter (WDM) and self-interacting dark matter (SIDM), modify the inner structure of haloes and predict different profile shapes and central slopes. We employ the AIDA-TNG simulations to investigate how alternative dark matter physics and baryonic processes jointly shape the internal structure of haloes. Using dark-matter-only and full-physics runs, we measure the dark matter density profiles of haloes spanning six orders of magnitude in mass, from  $10^{9.5} M_{\odot}$  to $10^{14.5}M_{\odot}$, considering radial bins that are well-resolved above the spatial resolution of the simulations ($r\geq4.4$ kpc and $r\geq1.7$ kpc for the 110.7 and 51.7 Mpc boxes, respectively). We fit the profiles with multiple analytical models and provide the distribution of the best-fitting parameters, as well as the concentration-mass relation in WDM and SIDM. The Einasto profile well reproduces the inner flattening produced in WDM models, both in the collisionless and in the full-physics runs. In SIDM dark-matter-only runs, haloes are better described by explicitly cored profiles, with core sizes that depend on mass and on the self-interaction model. When baryons are included, the differences between CDM and SIDM decrease, and such large dark-matter cores no longer form because adiabatic contraction in the baryon-dominated region counteracts self-interactions. Nevertheless, the coupling between baryons and self-interactions induces a broader range of inner slopes, including cases that are steeper than CDM at Milky Way masses. Alternative dark matter physics thus leaves clear signatures in the inner halo structure, even if baryons significantly reshape these differences. Our results are useful for future studies that need to predict the properties of haloes in multiple dark matter models.
}
   
   \maketitle


\section{Introduction}

The Cold Dark Matter (CDM) model assumes that dark matter consists of heavy, collisionless particles interacting only through gravity. In this framework, the formation and evolution of cosmic structures are well described by hierarchical clustering, and the internal structure of haloes arises naturally from gravitational collapse \citep{white78,blumenthal84,davis85}. In particular, the spherically averaged density profiles of dark matter haloes are well represented by the classic Navarro–Frenk–White (NFW) model \citep{navarro96,navarro97} or by the Einasto profile \citep{merritt06,navarro04,springel08b}. The compactness of the density distribution is commonly quantified through the concentration-mass relation. This relation has been extensively calibrated in collisionless simulations \citep{dolag04,duffy08,giocoli12b,ludlow13,ludlow16,dutton14}, showing that more massive haloes are less concentrated than less massive ones and that concentrations decrease with redshift \citep{bullock01a,wechsler02}. Hydrodynamical simulations have demonstrated that the introduction of baryonic processes significantly modifies this simple picture. The current generation of high-resolution hydrodynamical simulations \citep{schaye15,pillepich18b,nelson18,dave19} now models galaxy formation with sufficient realism to reproduce key properties of galaxies and their satellites over many orders of magnitude in mass. The processes involved, such as gas cooling, star formation, feedback, and dynamical interactions, can either steepen the central density profile through adiabatic contraction \citep{blumenthal86,gnedin04} or produce central cores through energetic feedback \citep{governato10,pontzen12}. As a result, the linear dependence between concentration and mass can be altered or even broken \citep{diemer15,lovell18,anbajagane22,sorini24}. 

The introduction of baryonic physics in simulations, together with a parallel improvement in the observational domain, has also contributed to solving some long-standing tensions between CDM predictions and observations, such as the missing satellite problem \citep{klypin99,drlica2015,kim18,engler21,wetzel16} or the formation of cored density profiles via feedback processes. However, other small-scale problems are still unsolved in CDM, as the diversity of galaxy rotation curves \citep{oman15,creasey17} and the abundance of over-concentrated systems detected via gravitational lensing \citep{meneghetti20,meneghetti23,minor2021,despali25b,enzi24}. Alternative dark matter models, such as warm (WDM) and self-interacting (SIDM) dark matter, are candidates to explain these discrepancies since they tend to modify the number density and density structure of haloes. The primary effect of WDM is to suppress the number of low-mass haloes compared to CDM, but the high velocity of particles naturally also lowers their concentration, producing a turnover in the concentration-mass relation \citep{lovell14,ludlow16,bose16}. In SIDM, non-gravitational interactions between particles lead to the exchange of energy and momentum in high-density regions and thus to the creation of a central density core, flatter than that seen in WDM and present at all halo masses \citep{vogel12,vogel16}. The size of this central core depends on the frequency of the interactions between dark matter particles, regulated by the self-interaction cross section: in dark-matter-only (DMO) simulations, a larger cross section corresponds to larger density cores. However, when the cross section is very high ($\sigma/m_{\chi}\geq$ 100 cm$^{2}$/g), a fast core formation can be followed by a second phase of gravothermal collapse towards the centre, leading to a rapid increase of the central density \citep{tran25,neev24,yang23,kong25}. 

Also in this case, baryonic effects complicate the picture by altering the timescales of core formation. An increasing number of recent works have looked at the combined effect of galaxy formation and alternative dark matter models. At the low-mass end, \citet{vogel14}, \citet{burger22} and \citet{gutcke25} simulated the evolution of dwarf galaxies in CDM and SIDM, combined with baryonic models \citep{Torrey14,gutcke21,gutcke22}. The TangoSIDM project \citep{correa22,correa24} instead simulated a 25 Mpc cosmological box in multiple SIDM models, aimed at the study of Milky-Way galaxies and their satellites. In the mass range of massive galaxies, \citet{despali19} and \citet{rose23} have shown that SIDM combined with the TNG galaxy formation model can lead to a variety of density profiles at the scales of Milky-Way galaxies and groups, including cored and cuspy objects. At cluster scale, the combined effects of baryons and self-interactions were explored in the BAHAMAS-SIDM runs \citep{robertson18b,harvey19} and recently in the DARKSKIES zoom-in simulations \citep{harvey25}. In this mass range, the self-interaction cross-section is constrained to be lower than $\sigma/m_{\chi}\geq0.5$ cm$^{2}$/g \citep{harvey15,kaplinghat16,sagunski21}. The diverse properties of dwarf galaxies are instead compatible with a scenario in which some systems are cored, and others are undergoing gravothermal collapse, requiring  $\sigma/m_{\chi}\sim 100$ cm$^{2}$/g \citep{correa21}. Therefore, in order to reconcile these two regimes, the cross-section should be velocity-dependent, such that particles moving in clusters of galaxies have small cross-sections, and particles in dwarfs have very high cross-sections.

While most simulation projects are interested in one specific mass range, here we aim at testing alternative dark matter models over a wide range of halo and galaxy types and environments. This is essential, not only to derive constraints and exclusion regions for alternative models, but also to provide scaling relations for WDM and SIDM that are as well calibrated  as their CDM counterparts. In this way, we can ask if any other model can explain the process of structure formation and match astrophysical observations as well as CDM. The AIDA-TNG simulations \citep{despali25a} are ideal for this purpose since they provide a consistent set of high-resolution cosmological boxes simulated with the same galaxy formation model and multiple WDM and SIDM models, in addition to CDM.  In this work, we characterise the dark matter profiles of CDM, SIDM, and WDM haloes spanning six orders of magnitude in mass, from $10^{9.5}$ to $10^{14.5}M_{\odot}$. We compare the DMO and full-physics (FP) runs to study the interplay between baryons and dark matter variations, breaking the degeneracy between the two at small scales.

The paper is structured as follows. We present the AIDA-TNG runs and the models for the density profiles in Sect. \ref{sec:sims} and Sect. \ref{sec:methods}, respectively. In Sect. \ref{sec:dmo}, we start by discussing the impact of alternative models on the profiles in the DMO runs and then move to the combined effect of these with baryonic physics in Sect. \ref{sec:hydro}, addressing the difference in adiabatic contraction between CDM, WDM, and SIDM. In both cases, we model the dark matter density distribution with multiple analytical models and derive the concentration-mass relation at $z=0$. In Sect. \ref{sec:redshift}, we discuss how the profile properties and best-fit parameters vary with redshift. Finally, we summarise the results and present our conclusions in Sect. \ref{sec:conc}. In the Appendices, we discuss more technical aspects related to the goodness-of-fit of different models (Appendix \ref{sec:app1}) and resolution effects (Appendix \ref{sec:app2}).


\section{The AIDA simulations} \label{sec:sims}

{\renewcommand{\arraystretch}{1.3}
\begin{table*}
\caption{Summary of the AIDA-TNG runs used in this work.}
\label{table:1}
\centering
\begin{tabular}{c c c c c c c| c c | c c }
\hline\hline
Name & Box & Physics & $m_{\rm DM}$ & $m_{\rm bar}$ & $\epsilon_{\rm DM,*}^{z=0}$ & CDM & WDM  & WDM  & SIDM  & vSIDM \\
& $\rm{[Mpc]}$&  & $ [\rm{M}_{\odot}$] & $ [\rm{M}_{\odot}$] & [kpc] & & 1\,keV & 3\,keV  & 1$\,\rm{cm^{2}g^{-1}}$ & Correa+21 \\
\hline\hline
   100/A& 110.7 & DMO & 7.1$\times 10^{7}$ & - & 1.48 &\ding{51} & - &\ding{51} &\ding{51}  &\ding{51} \\
   & & FP & 6.0$\times 10^{7}$ & 1.1$\times 10^{7}$ & 1.48 & \ding{51} & - & \ding{51}& \ding{51}  & \ding{51} \\
  100/B & 110.7 & DMO & 5.7$\times 10^{8}$ & -  & 2.95 &\color{lightgray}{\ding{51}} & \color{lightgray}{\ding{51}} & \color{lightgray}{\ding{51}} & \color{lightgray}{\ding{51}} & \color{lightgray}{\ding{51}} \\
    & & FP & 4.8$\times 10^{8}$ & 8.9$\times 10^{7}$ & 2.95& \color{lightgray}{\ding{51}} & \color{lightgray}{\ding{51}} & - & \color{lightgray}{\ding{51}}  &\color{lightgray}{\ding{51}} \\    
    \hline
   50/A &51.7 & DMO & 4.3$\times 10^{6}$ & - & 0.57 & \ding{51} & - & \ding{51}  & \ding{51}  & \ding{51} \\
   & & FP & 3.6$\times 10^{6}$ & 6.8$\times 10^{5}$ & 0.57 & \ding{51} & - &  \ding{51} &\ding{51}  &\ding{51} \\
   50/B & 51.7 & DMO & 3.4$\times 10^{7}$ & - & 1.15 & \ding{51} & \ding{51} & \color{lightgray}{\ding{51}} & \color{lightgray}{\ding{51}}  & \color{lightgray}{\ding{51}} \\
   & & FP & 2.9$\times 10^{7}$ & 5.4$\times 10^{6}$ & 1.15 & \ding{51} & \ding{51} & \color{lightgray}{\ding{51}} & \color{lightgray}{\ding{51}}  & \color{lightgray}{\ding{51}} \\

\hline\hline
\end{tabular}
\tablefoot{ We list the mass and spatial resolution of each run and mark the dark matter models that are used in this work. We use $\epsilon_{DM,*}$, the Plummer-equivalent softening length, to quantify the spatial resolution of each run. Those marked in grey are not used in the main body of the paper and only included for resolution convergence tests (see Appendix \ref{sec:app2}).}
\end{table*}
}

The AIDA-TNG simulations, \emph{Alternative Dark Matter in the TNG universe} (AIDA for short, hereafter), consist of cosmological volumes of increasing resolution and decreasing size simulated in cold and alternative dark matter models, with and without baryonic physics. In this work, we analyse the two largest boxes of 110.7 and 51.7 Mpc on a side, presented in \citet{despali25a}. The uniqueness of the AIDA set lies in the fact that we run the same cosmological volumes in multiple dark matter scenarios, spanning cold, warm, and self-interacting dark matter. Moreover, we create both the DMO and FP version of each run so that we can disentangle the effects of alternative dark matter and baryonic physics as described by the IllustrisTNG galaxy formation model \citep[TNG hereafter][]{weinberger17,pillepich18}. 

All simulations start at redshift $z=127$ and adopt the cosmological model from \citet{planckxxiv}, namely: $\Omega_{\rm m}=0.3089$, $\Omega_{\rm \Lambda}=0.6911$, $\Omega_{\rm b}=0.0486$, $H_{0}=0.6774$ and $\sigma_{8}=0.8159$. The 110.7 and 51.7 Mpc boxes (corresponding to 75 and 35 $h^{-1}{\rm Mpc}$, respectively) are run from the same initial conditions of the corresponding runs of the original IllustrisTNG simulations \citep{pillepich18b}. The AIDA sample includes five dark matter models, in addition to CDM:

\begin{itemize}
    \item Three WDM models with increasing particle mass: m$_{\rm WDM}=(1,3,5)\,{\rm keV}$, referred to as WDM1, WDM3, and WDM5 (the latter not used in this work).
    \item Two self-interacting dark matter models: SIDM1 is a classic elastic SIDM model with constant cross-section $\sigma/m_{\chi}=1\,{\rm cm^{2}g^{-1}}$, while in vSIDM the interaction cross-section is velocity-dependent and follows the model by \citet{correa21}.
\end{itemize}
Table \ref{table:1} summarises the properties of the runs used in this work: we consider the high-resolution boxes 50/A and 100/A for all dark matter models, except for WDM1, where the 50/B run is used. Since the WDM1 model already starts to suppress structures at the scale of $\sim10^{10}M_{\odot}$, the higher-resolution 50/A simulation was not created since it would not have added new information. We refer the reader to the presentation paper \citep{despali25a} for a more detailed description of the simulations and the considered models.

Haloes and subhaloes are identified using the \textsc{SUBFIND} algorithm, producing catalogues of the halo and galaxy properties with the same information available for the IllustrisTNG runs \citep{nelson19}. Throughout the paper, we define halo masses as $M_{200c}$, corresponding to the mass inside the radius  $R_{200c}$  where the enclosed density equals 200 times the critical density at the redshift of the considered snapshot. Masses and lengths are given in units of $M_{\odot}$ and comoving kpc, respectively.

\section{Methods} \label{sec:methods}
\subsection{Simulated profiles}
\begin{figure*}
\centering
    \includegraphics[width=\textwidth]{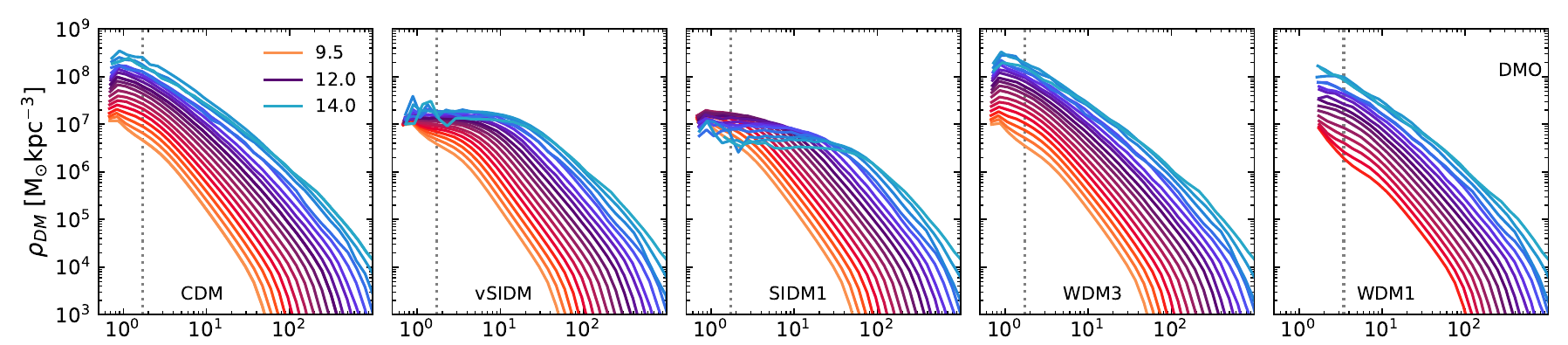}
    \includegraphics[width=\textwidth]{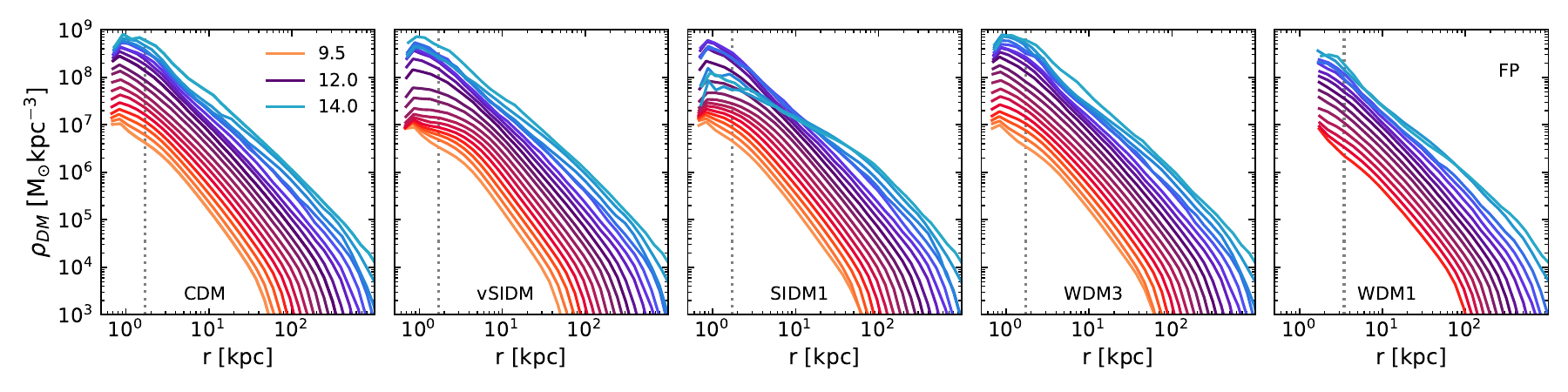}
    \includegraphics[width=0.97\textwidth]{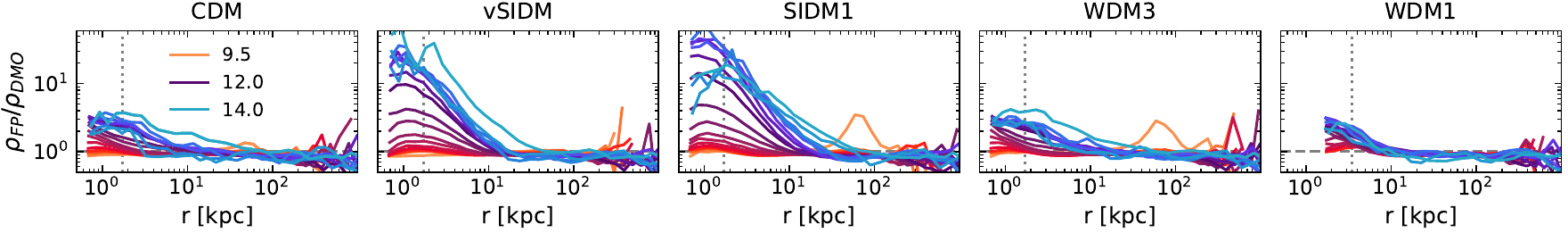}

    \caption{Mean dark matter profiles for haloes in 21 logarithmically-spaced bins in mass between $3\times10^{9}M_{\odot}$ (orange) and $10^{14}M_{\odot}$ (light blue), from the DMO (top) and FP runs (middle). From left to right, we show CDM, followed by two self-interacting models (vSIDM and SIDM1) and two WDM models (WDM3 and WDM1) . We use the 50/A run in the first four panels and the WDM1 50/B box in the rightmost one. The latter only allows us to reliably measure density profiles of $M_{200c}\geq10^{10.5}M_{\odot}$, with stronger limits on the spatial resolution. In all panels, the dashed vertical line stands for the resolution limit $3\epsilon_{DM}$. When fitting the density profile, we restrict the range further to $r\geq3\epsilon_{DM}$ and haloes with more than 1000 dark matter particles. 
        Finally, the bottom panels highlights the contraction caused by baryons by showing the ratio between the dark matter profiles in the FP and DMO cases.
    } \label{fig:prof}
\end{figure*}

\begin{figure*}

    \includegraphics[width=\textwidth]{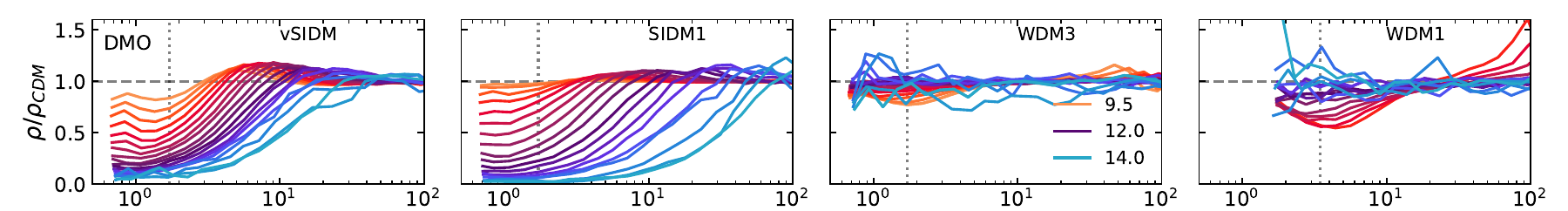}
    \includegraphics[width=\textwidth]{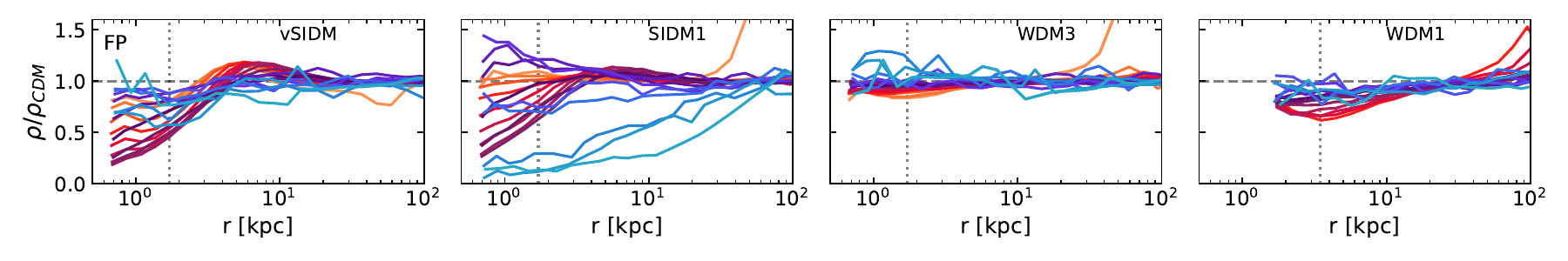}

    \caption{Using the same mass bins of Fig. \ref{fig:prof}, we show the ratio between the dark matter density in each alternative scenario and the CDM one. The top and bottom panels show the result for the DMO and FP runs, respectively. } \label{fig:prof_res}
\end{figure*}

The AIDA boxes allow us to explore many orders of magnitude in halo mass and, at the same time, to measure the density distribution inside haloes down to kpc scales. The 51.7 Mpc boxes are essential to reliably measure the density profiles of haloes with $M_{\rm 200c}\leq10^{11}\,{\rm M}_{\odot}$, due to the higher mass and spatial resolution. At the same time, the larger 110.7 Mpc boxes provide us with the best statistical sample, especially at the high-mass end, $M_{\rm 200c}\geq10^{12.5}\,{\rm M}_{\odot}$. We measure the density profiles of each halo by binning particles in 40 logarithmically spaced bins in the range between 0.4 kpc and the virial radius of the halo. The density within each radial bin is then computed as the ratio of the total mass of dark matter particles falling within said bin and the enclosed spherical volume. 
As an example, Fig. \ref{fig:prof} shows the mean dark matter density profiles of haloes in 21 logarithmically-spaced mass bins, with a bin width of 0.2 dex. The top panels show the profiles in the DMO runs, while the bottom panels focus on the dark matter profiles of the FP case, where we can appreciate the dark matter back-reaction to the presence of baryons. We can see that in the DMO runs (top panels), self-interactions create wide flat cores in the dark matter distribution, resulting in larger cores for higher masses. On the other hand, WDM reduces the central density in the lower mass bins, leaving more massive haloes unchanged. While DMO results appear qualitatively clean and regular, the addition of baryons in the FP runs (middle panels) complicates the picture, creating different trends in vSIDM and SIDM1. Figure \ref{fig:prof_res} highlights the mass and radial dependence of the differences between dark matter models by showing the ratio of all profiles to the CDM case.

The combination of the two simulation volumes allows us to benefit from the high resolution of the 50/A runs while also exploiting the larger statistical sample provided by the 100/A boxes. However, the differing mass and spatial resolutions must be considered carefully, as they can introduce systematic effects in the inner regions of halo profiles. In particular, caution is required when comparing FP runs across resolutions. In Appendix \ref{sec:app2}, we present a set of convergence tests and discuss the origin of the differences observed among the dark matter models. In summary, we find good convergence in the dark-matter density profiles beyond our adopted lower limit of  $r=3\epsilon_{\rm DM}$  for CDM, WDM, and for the DMO versions of SIDM1 and vSIDM. When baryons are included in the self-interacting runs, the stellar distributions and the resulting dark-matter response show a resolution- and mass-dependent behaviour, with the largest discrepancies occurring for haloes with $M_{\rm 200c}=[10^{12},\,10^{13}]M_{\odot}$, where the stellar component plays a dominant role. For these reasons, we adopt the 50/A simulations as our fiducial description of halo profiles, and discuss how they can be consistently combined with the lower-resolution runs.

\subsection{Analytical models}

In \citet{despali25a}, we considered the NFW \citep{navarro96} model to describe the density profiles and showed the concentration-mass relation at $z=0$ based on its best-fit parameters. Given the variety of inner slopes of the density profile created by WDM and SIDM, here we test two additional analytical models that better describe the modifications introduced by alternative dark matter scenarios.

The first is the Einasto profile \citep{einasto65} defined as:
\begin{equation}
\label{Einasto}
\rho(r) = \rho_{\text{E}}\,\, \exp \left\{-\frac{2}{\alpha_{\text{E}}}\left[\left(\frac{r}{r_{\text{E}}}\right)^{\alpha_{\text{E}}}-1\right]\right\},
\end{equation}
where $r_{\text{E}}$ is the radius at which the logarithmic slope equals $-2$, $\rho_{\text{E}}$ is the corresponding density,  and $\alpha_{\text{E}}$ controls the curvature of the profile. The additional shape parameter $\alpha_{\text{E}}$ makes the Einasto profile more flexible than NFW and allows it to reproduce both cuspy and cored density distributions. Several studies have demonstrated that the Einasto form fits CDM haloes at least as well as, and often better than, the NFW profile \citep[e.g.][]{merritt06,gao08,retana12,despali20}, with typical values $\alpha_{\text{E}} \approx 0.2$ across a wide range of halo masses \citep{springel08b,ludlow16}.

Concentration can be defined analogously for NFW and Einasto profiles, as the ratio between the $r_{\rm 200c}$ and the scale radius \citep{meneghetti14}. The concentration–mass relation of \citet{ludlow16}, for instance, is calibrated on Einasto fits to CDM and WDM haloes, yet yields concentration values compatible with NFW-based modelling such as those of \citet{sorini24}. 

The second model we consider is the parametric function introduced by \citet{gilman21} and \citet{yang24} (hereafter “Y24”), specifically designed to capture the characteristic features of SIDM haloes, including both central cores and the potential central density increase associated with gravothermal core collapse. It is defined as
\begin{equation}
    \rho_{Y}(r)=\frac{\rho_{Y}}{[(r^{\beta}+r^{\beta}_{c})^{1/\beta}/r_{Y}]\left(1+\frac{r}{r_{Y}}\right)^{2}}, \label{eq:yang}
\end{equation}
where $\rho_{Y}$ and $r_{Y}$ are the scale density and radius, and $r_{c}$ is the core size. The transition between the inner core and outer NFW profile is controlled by $\beta$. \citet{yang24} showed that $\beta=4$ is a good fit to simulated data.
In the limiting case of no core ($r_{c}=0$), this model reduces exactly to the NFW profile. Also in this case, we use the scale radius $r_{Y}$ to calculate the halo concentration. A comparable approach was adopted by \citet{tran24}, who also propose a density profile designed for SIDM haloes, yielding similar results, although with a better agreement for velocity-dispersion profiles. 

\begin{figure*}
\centering
    \includegraphics[width=0.9\textwidth]{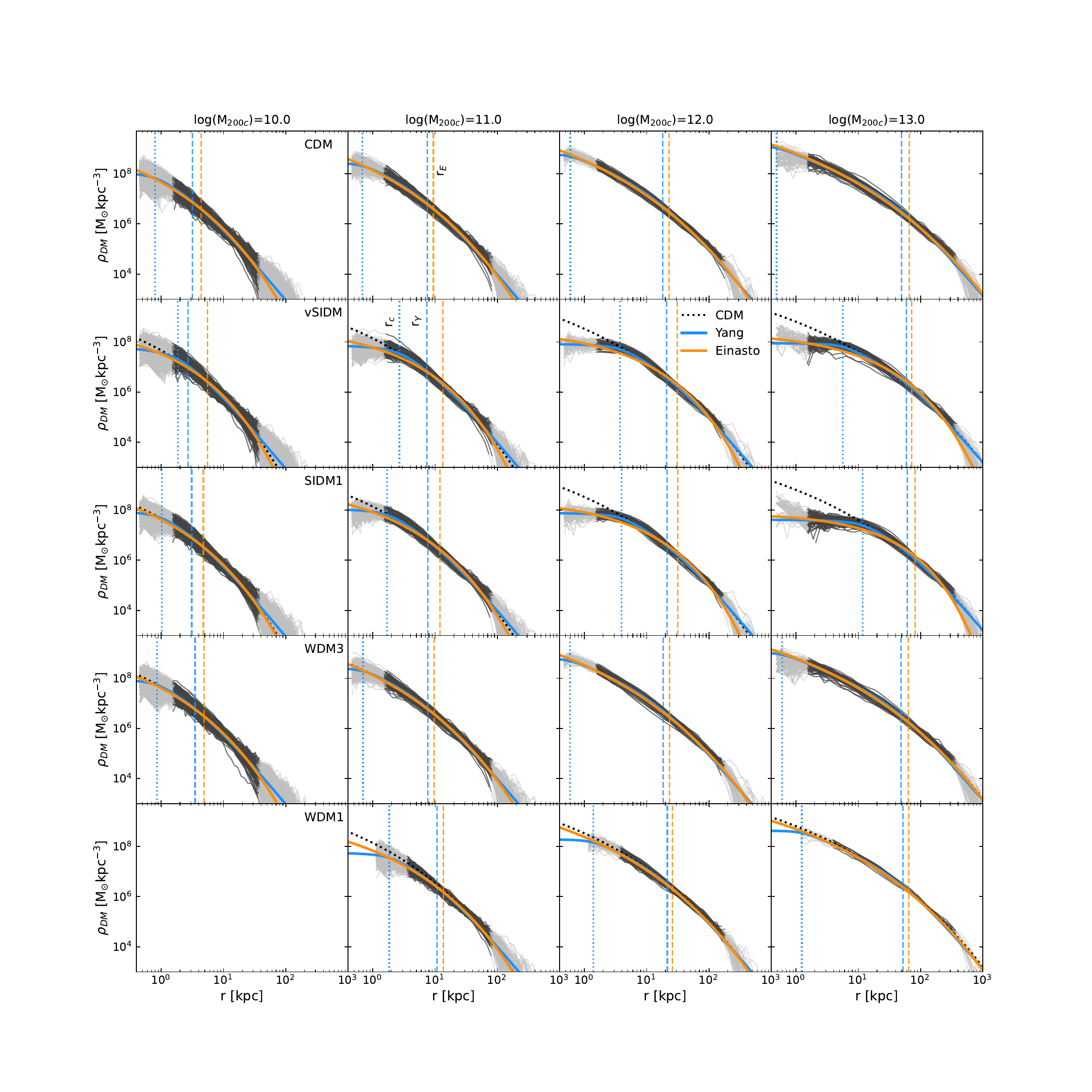}
    \caption{Examples of halo profiles and best fit in four bins of increasing mass at $z=0$ in the DMO runs. We show a different dark matter model in each row. In each panel the grey lines show individual density profiles of haloes in the considered mass bin, while the black line indicates the radial range used to perform the fit ($3\epsilon_{DM}\leq r\leq r_{200c}$). The Einasto and Y24 best fits to the mean dark matter profile are shown in orange and blue. In the alternative dark matter panels, we also plot the CDM mean as the dotted black line for comparison. The vertical dashed lines of corresponding colour mark the two scale radii $r_{E}$ and $r_{Y}$, while the vertical dotted blue line stands for the Y24 core radius $r_{c}$. As we only considered systems with more than 1000 particles, we do not show the profiles of WDM1 haloes in the first mass bin. } \label{fig:prof_bins}
\end{figure*}

We fit the dark matter density profiles with all analytical models, independently of the underlying dark matter physics, in order to evaluate which parametrisation best describes CDM, WDM, and SIDM haloes. To ensure the robustness of the fits, we restrict the analysis to radial bins in the range ($3\epsilon_{\rm DM} \le r \le r_{200c}$) (see Fig. \ref{fig:prof_bins}), where $\epsilon_{\rm DM}$ is the Plummer-equivalent softening length; the lower limit excludes the innermost region affected by numerical two-body relaxation and force softening.  This means that we consider $M_{\rm 200c}\geq5\times10^{9}M_{\odot}$ and $r\geq1.7$ kpc in the 50/A boxes, while $M_{\rm 200c}\geq7\times10^{10}M_{\odot}$ and $r\geq4.4$ kpc in the 100/A and 50/B ones.

\section{Density profiles in the dark-matter-only runs} \label{sec:dmo}
In this section, we model the effect of WDM and SIDM on the dark matter profiles in the AIDA DMO runs. This serves as a first baseline to understand the effects of dark matter physics without other intervening elements. We will then consider the FP runs in Sect. \ref{sec:hydro}. 

\begin{figure*}
    \centering
    \includegraphics[width=0.96\textwidth]{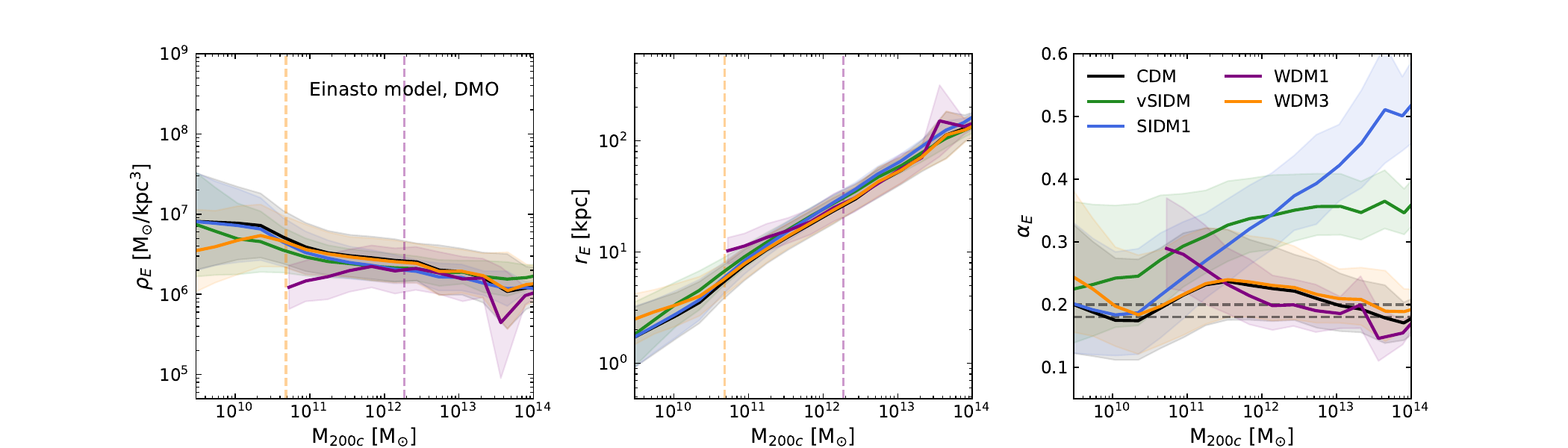}
    \includegraphics[width=0.96\textwidth]{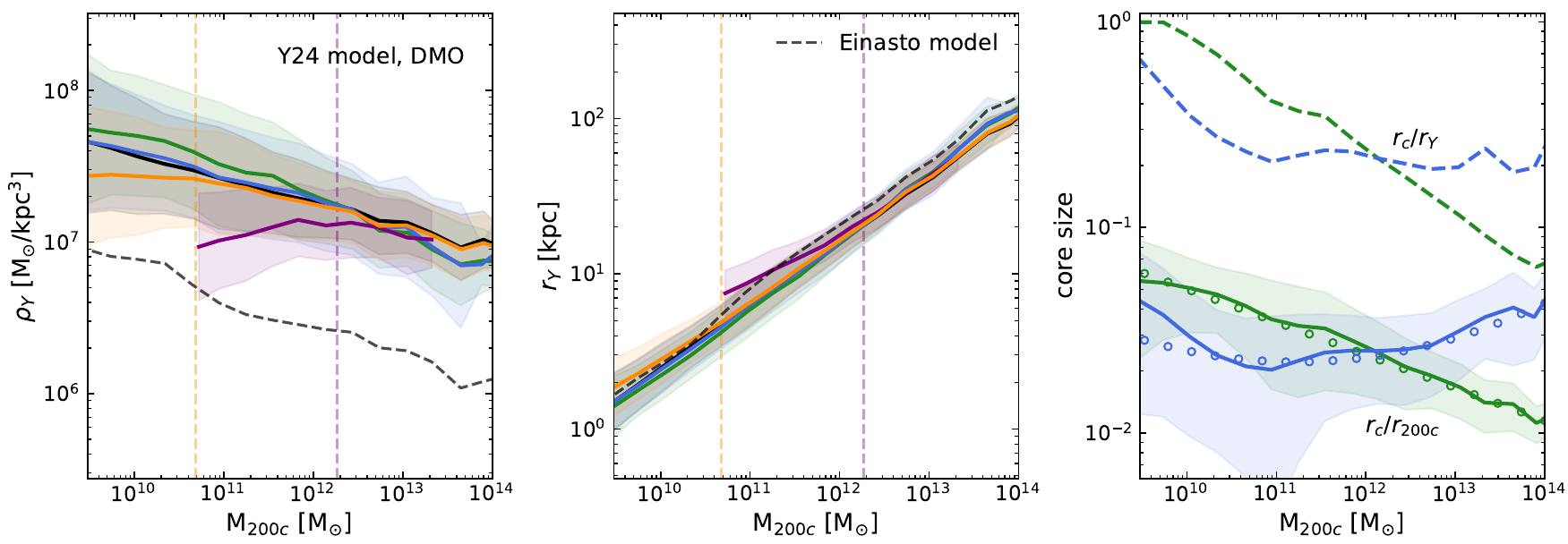}
    \includegraphics[width=0.42\textwidth]{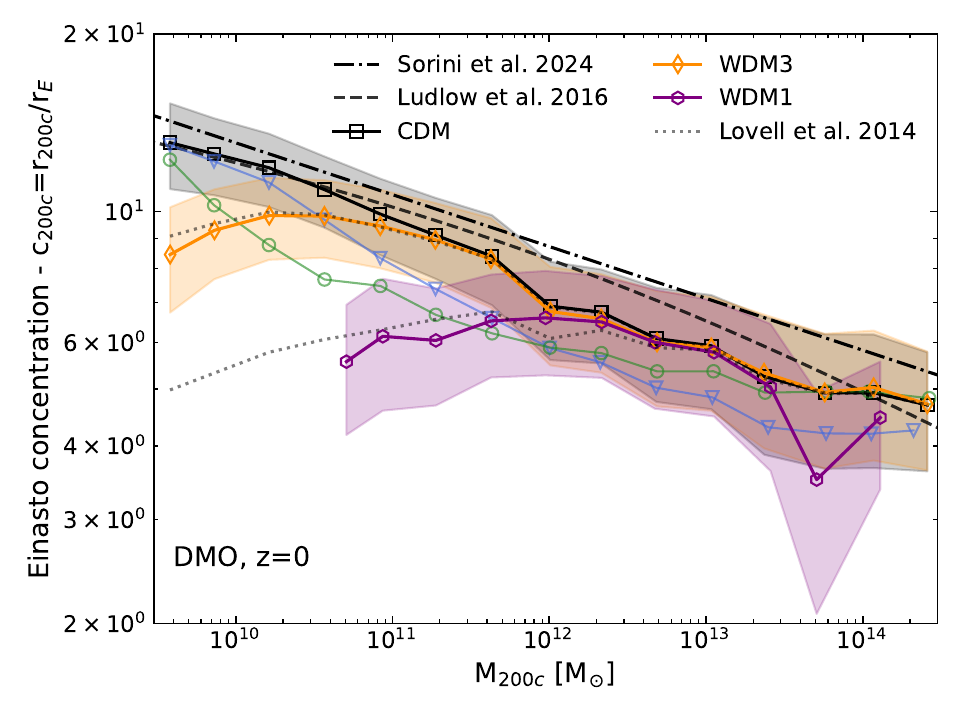}
    \includegraphics[width=0.42\textwidth]{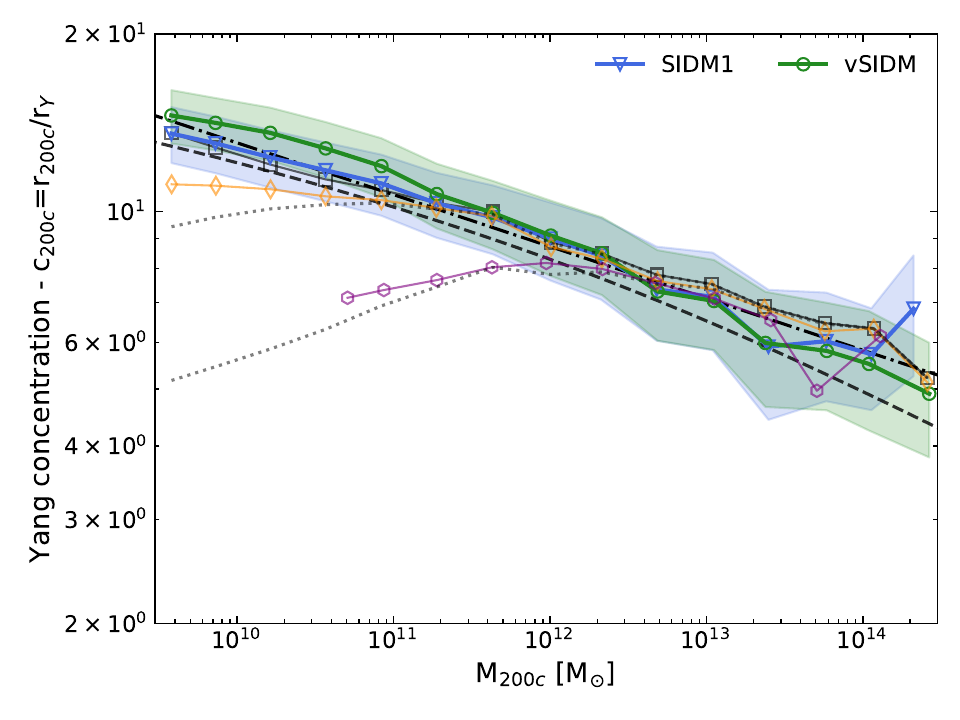}

    \caption{$Top$: Mean best fit parameters of the Einasto (first row) and Y24 (second row) models, plotted as a function of halo mass at $z=0$ in the DMO runs, together with their $1\sigma$ uncertainties represented by the coloured bands. In both cases, the first two panels show the mean scale density and radius, while the third panels differ: in the case of Einasto we have the shape parameter $\alpha_{E}$ (compared to the fixed range 0.18-0.2 used in previous works for CDM haloes), while in the Y24 model the core radius that we scale to the halo radius as $r_{c}/r_{200c}$ (solid lines) and to the scale radius (dashed lines). In the first two panels of the second row, the dashed black lines show the scale density and radius calculated from the Einasto fit in CDM, for comparison. Instead, in the right panel, the core radius is consistent with zero for the CDM and WDM models, and it is thus only shown for self-interacting scenarios. Here, the open circles correspond to the best-fit models from Eq. \ref{eq:model}. Finally, the vertical dashed lines correspond to the value 100$M_{hm}$, calculated for WDM1 and WDM3, where it is expected that the halo concentration starts deviating from CDM. $ Bottom$: Concentration-mass relation at $z=0$ for the DMO runs. On the left (right), we show the results based on the Einasto (Y24) fit. We compare to \citet{ludlow16}, who also used Einasto profiles to fit CDM and WDM halo profiles, and to \citet{sorini24}, who instead modelled haloes with NFW. The dotted line shows the predicted WDM concentration when applying the correction by \citet{lovell14} to the measured CDM relation (black squares). In the previous section, we have demonstrated that CDM and WDM profiles are best fit by Einasto, while SIDM ones by the Y24 profiles. We highlight these with solid lines in the figure and display the remaining ones with lighter colours.}
    \label{fig:ein_mean0}
\end{figure*}

\subsection{Dark matter profiles at $z=0$}

As described in Sect. \ref{sec:methods}, we measure dark matter density profiles in the AIDA simulations and model them using three analytical parametrisations: NFW, Einasto, and Y24. We fit both individual halo profiles and the mean profile in each mass bin. As expected, the NFW profile provides a good description of CDM haloes but lacks the flexibility to reproduce the cores induced by WDM and SIDM. Consequently, NFW concentrations are lower at the low-mass (high-mass) end for WDM (SIDM) haloes, in agreement with \citet{despali25a}.  Figure \ref{fig:prof_bins} shows dark matter profiles for haloes in four representative mass bins, from $M_{200c} = 10^{10} M_{\odot}$ to $M_{200c} = 10^{13} M_{\odot}$ from left to right. Individual halo profiles are plotted in light grey, while the dark grey part indicates the radial range used for the fit, according to the criteria described in the previous section. The orange and light blue curves show instead the best-fit Einasto and Y24 models to the mean profile in each bin. Overall, we find that the Einasto profile provides a reasonable description for all dark matter models, although it cannot fully capture the flattened cores present in  SIDM1 and vSIDM haloes, particularly for $M_{200c} \gtrsim 10^{12}M_{\odot}$. In contrast, the Y24 model well reproduces the central cores of SIDM haloes but does not accommodate the shallower WDM cores, which are better described by Einasto. We also note that the Y24 core radius $r_c$ (dotted vertical line) is always smaller than the scale radius $r_Y$ (dashed line), indicating that the core never extends to the region where the profile slope reaches -2. Appendix \ref{sec:app1} presents a more detailed analysis of the goodness-of-fit distributions for each model and dark matter scenario. Given that the Einasto and NFW profiles perform equally well for CDM, we adopt the Einasto profile for all subsequent analyses and do not discuss NFW further. Moreover, in the absence of a significant core, the Y24 model reduces to NFW, effectively including NFW as a potential model.

We now move to studying the distribution of the best-fit parameters of the two profiles: for this, we fit individual halo profiles and calculate the running means of scale density, scale radius and the additional parameters ($\alpha_{E}$ or $r_{c}$) regulating the slope transition. The results are shown in the first two rows of Fig.~\ref{fig:ein_mean0}. The scale densities and scale radii of both models follow similar trends across all dark matter scenarios, with small deviations at the low-mass end for WDM haloes. This behaviour is consistent with expectations, as WDM concentrations decrease relative to CDM below masses roughly 100 times the model half-mode mass, indicated by the vertical dashed lines. The largest differences among dark matter models appear in the third parameter of each profile. For Einasto, the shape parameter $\alpha_{\rm E}$ remains approximately constant with halo mass for CDM and WDM3, in agreement within 1$\sigma$ with the fixed value of $\alpha_{E}=0.18-0.2$ used in previous studies \citep{springel08b,mastromarino23}. WDM1 exhibits a small increase at the low-mass end, consistently with the deviations in the other two parameters of the fit and the cores forming at the low-mass end. In contrast, self-interacting models exhibit a clear mass dependence of $\alpha_{E}$, reflecting the increasing size of the central core, with a steeper trend for SIDM1 than for vSIDM. This trend is mirrored in the Y24 core radius $r_{\rm c}$, shown in the right middle panel: for high-mass haloes, the core is larger in SIDM1 than in vSIDM, whereas the trend reverses at low masses. This behaviour is consistent with the velocity-dependent self-interaction cross-section in vSIDM: particles in low-mass systems experience larger effective cross-sections, while high-velocity particles in massive haloes interact less. In the vSIDM model, the cross-section reaches $\sigma/m_{\chi} \sim$ 1 cm$^{2}$g$^{-1}$ for relative particle velocities $<v>\sim300$ km/s, corresponding to halo masses of $M_{200c} \sim 10^{13} M_{\odot}$ not far from the scale where the mean values of $\alpha_{E}$ and r$_{c}$ intersect. The correspondence is not perfect, given that vSIDM haloes experience different levels of interaction during their growth history, leading to an integrated effect.

\subsection{Concentration-mass relation}  \label{subsec:cm} 

We use the best-fit models derived in the previous section to measure halo concentrations, defined as $c_{200c}$=$r_{200c}/r_{s}$, where $r_{s}$ denotes the scale radius of the adopted profile (i.e. $r_{E}$ for Einasto and $r_{Y}$ for Y24). Given that both scale radii exhibit a similar linear dependence on mass (see Fig. \ref{fig:ein_mean0}), the resulting concentration–mass relations are expected to follow analogous trends. 
In the left panel, the WDM1 and WDM3 Einasto concentrations converge to the CDM prediction at the high-mass end, while exhibiting the characteristic downturn for low-mass haloes. The behaviour agrees well with the expectations from previous works \citep{lovell14,ludlow16}, where the mass scale at which suppression starts to be significant is a function of the WDM half-mode mass. In contrast, the SIDM1 and vSIDM Einasto concentrations remain systematically lower than CDM in all masses. This is a direct consequence of the inner core, whose presence is not fully absorbed by the Einasto scale radius.

The situation is different for the Y24 profile (right panel). Here, the inferred concentrations for both SIDM models closely match the CDM relation. This arises naturally from the fact that r$_{c}\leq$r$_{Y}$ and thus the scale radius of the Y24 model remains insensitive to core formation. The concentration-mass relation is thus an incomplete description for SIDM profiles. A more precise parametrisation can be obtained by combining the concentration–mass relation with the mass dependence of the core size, shown Fig. \ref{fig:ein_mean0}. Setting $x=\log{M_{200c}}$ and $y=\log(r_{c}/r_{200c}$), we derive:  
\begin{eqnarray}
    y\text{(SIDM1)}=0.37-0.5x+0.0026x^2 \qquad \text{and} \nonumber\\
    y\text{(vSIDM)} = 0.16-0.11x . \label{eq:model}
\end{eqnarray}
The resulting predictions are shown as open circles in the right panel of Fig.~\ref{fig:ein_mean0}. As we discuss in the next section, the inclusion of baryons further complicates this picture.

\section{Modelling the effect of baryons on dark matter profiles} \label{sec:hydro}

In this section, we analyse the dark matter density profiles in the FP runs to quantify how baryonic processes reshape halo structure across mass and dark matter models. The inclusion of gas cooling, star formation, and feedback can either steepen or flatten the inner density profile, depending on the interplay between baryonic contraction and energy injection. These competing effects are expected to introduce mass- and model-dependent modifications, particularly in haloes where the stellar component contributes significantly to the central gravitational potential. By comparing FP profiles to their DMO counterparts, we aim to identify the regimes in which baryons enhance or erase the distinct features of each dark matter scenario—such as the WDM suppression, SIDM cores, or the Einasto-like curvature in CDM haloes—and to assess how robust these signatures remain once galaxy formation is taken into account.

\subsection{Adiabatic contraction in SIDM} \label{subsec:adiabatic}

\begin{figure*}
    \centering
    \includegraphics[width=\linewidth]{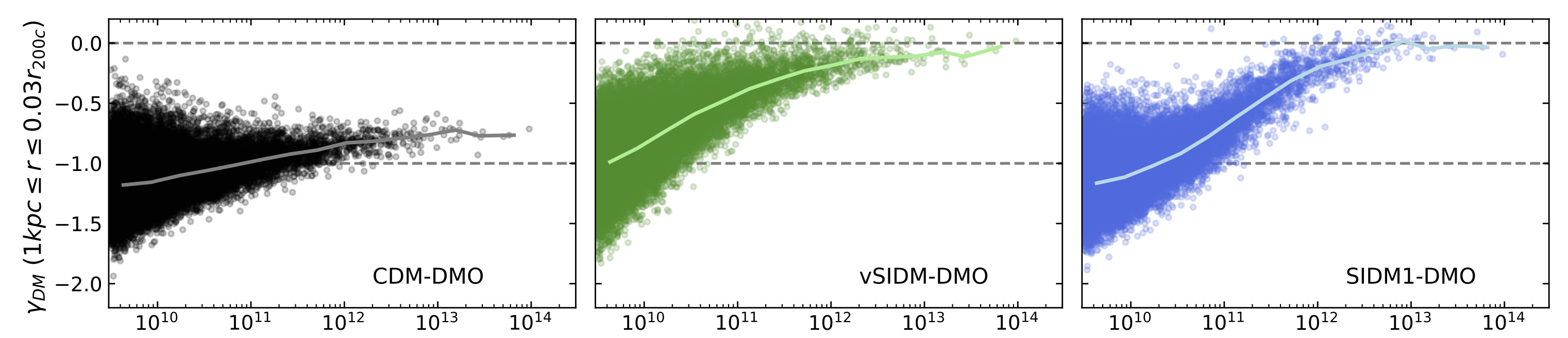}
    \includegraphics[width=\linewidth]{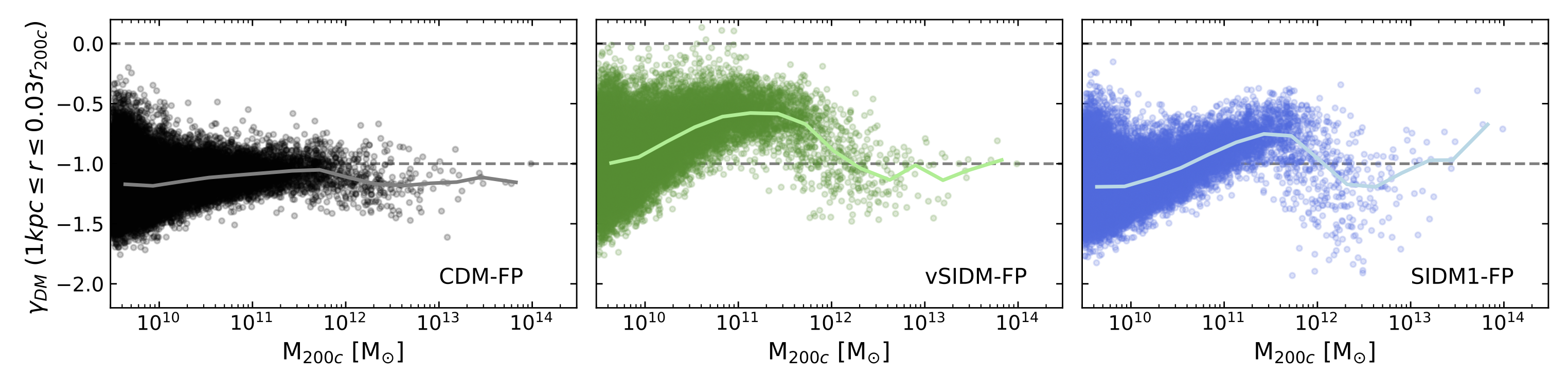}
    \caption{Slope of the inner dark matter profile $\gamma_{DM}$ at $z=0$, comparing CDM haloes (black) to SIDM1 (blue) and vSIDM (green), in the DMO (top) and FP (bottom) runs. We measure the slope of the profile in the radial range 1 kpc$\leq r\leq0.03r_{200c}$, where the latter is chosen based on the Y24 core size distribution from Fig.\ref{fig:ein_mean0}. For low-mass haloes, it may happen that this value is too close to the resolution limit, and thus in those cases we choose $r$=10 kpc as an upper limit. The solid lines show the running medians for each distribution.}
    \label{fig:prof_slope}
\end{figure*}

\begin{figure*}
    \centering
    \includegraphics[width=\linewidth]{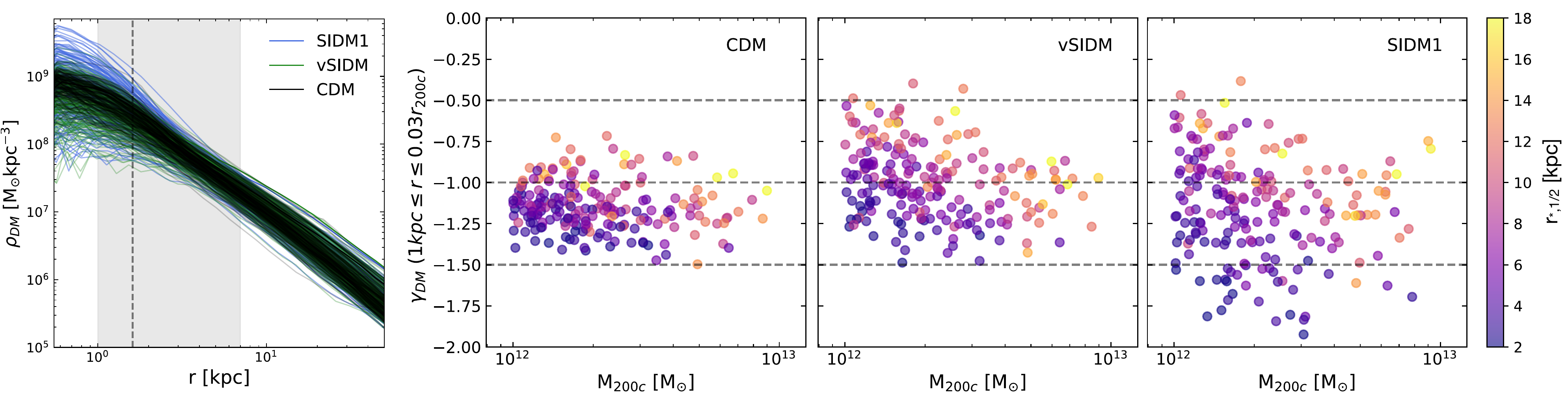}
    \caption{\emph{Left}: dark matter density profiles of haloes in the mass range $\log M_{200c}=[10^{12}-10^{13}]$ $M_{\odot}$ in the 50/A run, for the CDM (black), vSIDM (green) and SIDM1 (blue) FP runs. We only show the profiles for $r\geq\epsilon_{DM}$ and the vertical dashed line marks $r=3\epsilon_{DM}$, while the grey band shows is the range used to measure $\gamma_{DM}$. \emph{Right}: Inner slope of the dark matter density profiles from the left panel, plotted as a function of halo mass for the three runs and coloured by the stellar half-mass radius r$_{*,1/2}$ of the central galaxy. }
    \label{fig:prof_m12}
\end{figure*}

It is well established from CDM simulations that the presence of baryons in the central regions of haloes leads to an increase in the dark matter density through adiabatic contraction: as baryons cool and collapse to form the central galaxy, the deepening of the gravitational potential causes the dark matter to move inward, steepening the inner density profile. This behaviour is evident in Fig. \ref{fig:prof}, where all FP runs show an enhancement of the central dark matter density relative to their DMO counterparts. The effect is strongest at the high-mass end, where massive central galaxies produce a pronounced bump in the inner profile. At lower masses, stellar and supernova feedback partly counteract the contraction by injecting energy into the central regions and temporarily flattening the inner profile. Given that self-interactions create flat large cores in the DMO runs, it is worth discussing separately what is the effect of baryonic contraction in this case, before moving to the main results on density profile models.

In the bottom panels of Fig. \ref{fig:prof}, we have seen the ratio between the FP and DMO dark matter profiles in the same mass: here it was already clear that the effect of adiabatic contraction is stronger in self-interacting scenarios, where the density increases to a factor of $\sim$ 30 in the centre at the high-mass end, compared to a factor of $\sim$4 in CDM and WDM. In the mass range where stars dominate the inner profile, the additional gravitational pull created by the central galaxy prevents the formation of the large core present in the DMO runs. As a result, the distribution of inner slopes is strongly modified. Figure \ref{fig:prof_slope} compares the inner density slopes measured in the DMO (top) and FP (bottom) simulations. In the DMO case, CDM haloes cluster around the NFW expectation  ($\gamma_{DM}=-1$), while SIDM1 and vSIDM haloes develop shallower slopes that approach fully cored profiles ($\gamma_{DM}=0$) at high mass. Once baryons are included, all profiles steepen and the flattening induced by self-interactions is significantly reduced: in the FP case, only a few systems retain slopes shallower than $\gamma_{DM}=-0.5$ and the mean value turns around at $M_{200c}\sim4\times10^{11}M_{\odot}$, where the stellar mass fraction starts to be significant. Both SIDM models still exhibit a broader diversity of inner slopes than CDM, but the allowed range is narrower than in the DMO case, and the mass dependence becomes more intricate.

We also identify a subset of SIDM1 haloes in the mass range $M_{200c}\sim10^{12-13}M_{\odot}$ that develop steep cusps with $\gamma<-1.5$. This behaviour is consistent with previous studies that combined SIDM with the TNG galaxy formation model \citep{despali19,rose23}, which have demonstrated that in this mass range the interplay between baryons and dark matter can yield a wide variety of inner slopes. In particular, \citet{rose23} showed that the formation of a compact central galaxy increases the velocity dispersion in the inner halo relative to the DMO case, making the centre the hottest region. SIDM scatterings then transport heat away from the centre, removing thermal support and driving a rise in central density—a process analogous to gravothermal contraction, where the system behaves as if it had a negative specific heat. This mechanism represents the precursor stage of gravothermal collapse. This diversity is illustrated in Fig. \ref{fig:prof_m12}. The left panel shows that SIDM1 haloes in the  $M_{200c}\sim10^{12-13}M_{\odot}$ mass interval span a wider range of inner slopes, including profiles significantly steeper than those in CDM or vSIDM. The right panels explore the connection between the inner dark matter slope and the compactness of the central galaxy, measured via the stellar half-mass radius $r_{*,1/2}$ and shown by the colour scale. Systems hosting compact stellar components tend to have the steepest SIDM1 profiles, consistent with the correlations identified by \citet{despali19} between dark matter contraction, galaxy size, and formation time. The larger statistical sample of AIDA compared to previous works enables a more comprehensive exploration of these processes; we plan to follow the redshift evolution of individual systems using merger trees in upcoming work.

Finally, we note that the stellar profiles themselves show no significant differences between dark matter models. This indicates that the enhanced FP–DMO discrepancy in SIDM haloes arises primarily from the dark matter response to the baryonic potential, rather than from variations in the baryonic distribution across scenarios. As discussed in Sect. \ref{sec:methods}, the baryonic content of haloes — and therefore the degree of contraction—depends on resolution. In Appendix \ref{sec:app2}, we present convergence tests for both dark matter and stellar profiles across the four resolution levels of AIDA. We also note that the AIDA-TNG haloes do not form density cores due to feedback. It has been shown in previous hydrodynamical simulations \citep{pontzen12} that bursty stellar feedback can induce rapid fluctuations in the central gravitational potential, transferring orbital energy to dark matter particles and this leading to flattening of NFW-like cusps into kpc-scale cores in dwarf galaxies ($10^{10}$ - $10^{11}M_{\odot}$). This process depends on resolution and on the details of the galaxy formation implementation. In AIDA, but also in the IllustrisTNG simulations, we do not observe the formation of significant density cores in the FP run compared to its DMO counterpart.

\begin{figure*}
    \centering
    \includegraphics[width=\textwidth]{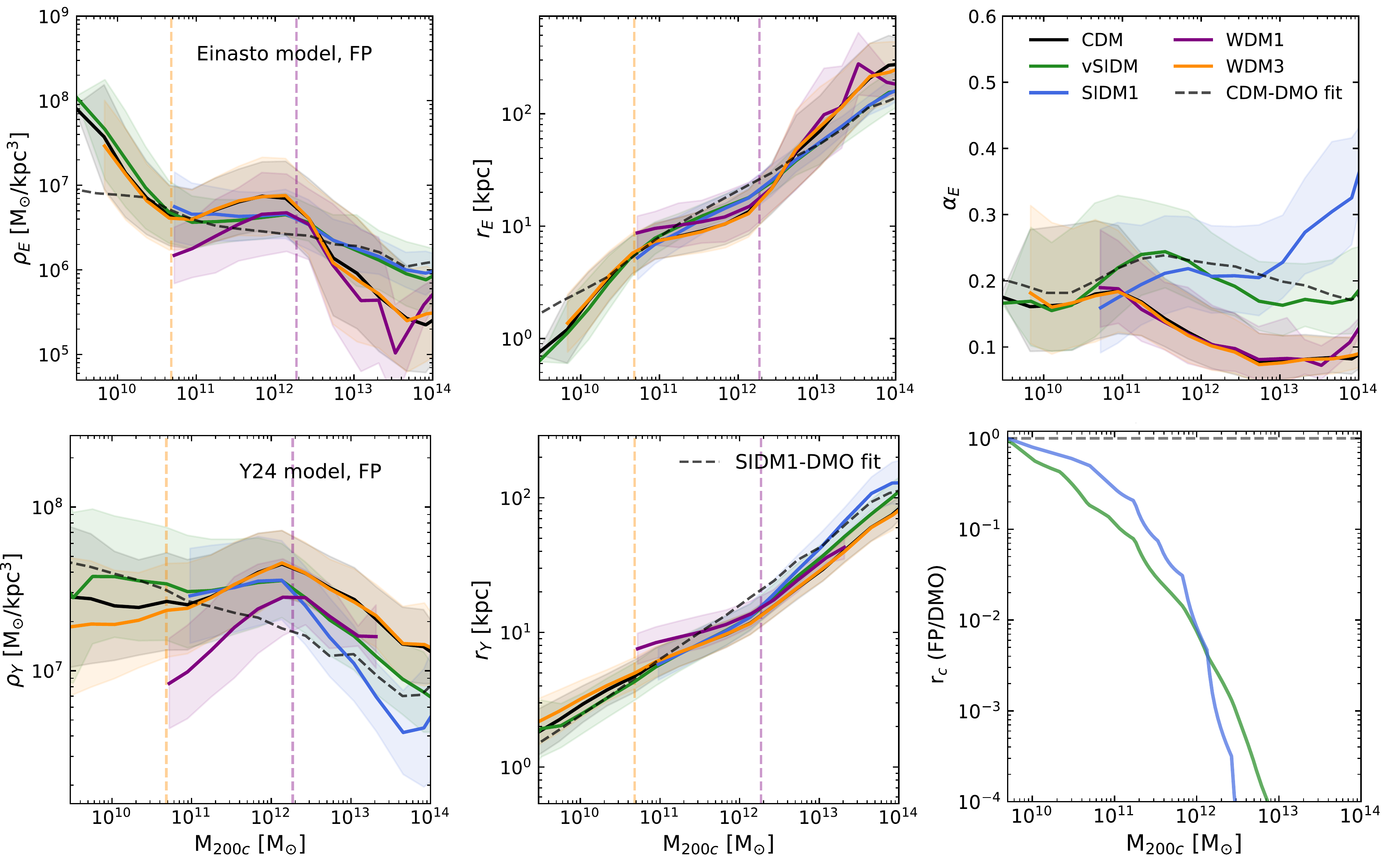}
    \includegraphics[width=0.45\textwidth]{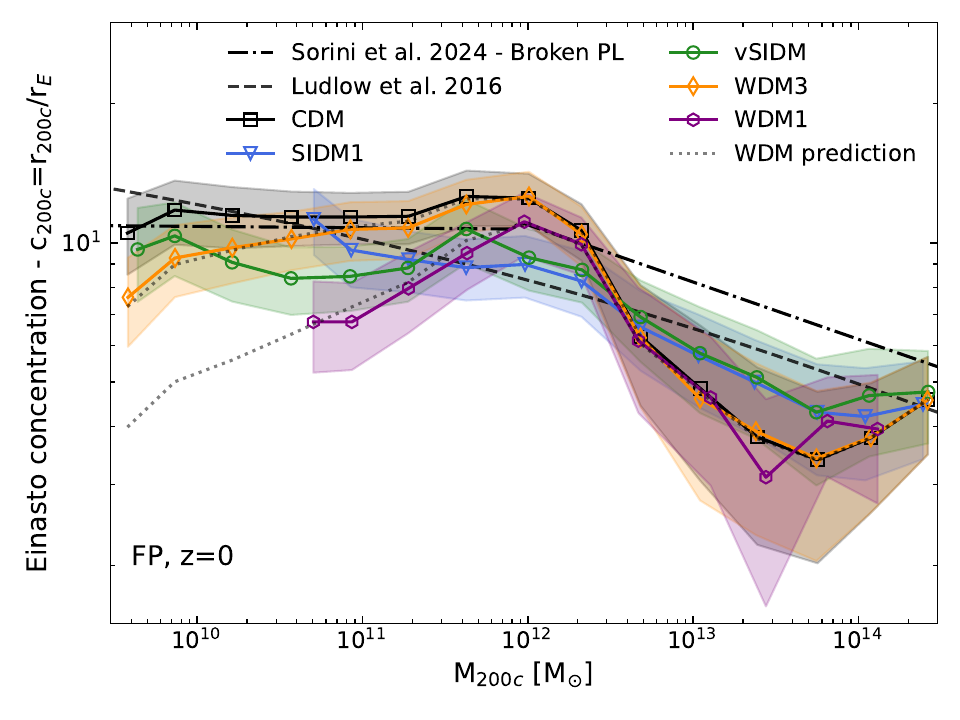}
    \includegraphics[width=0.45\textwidth]{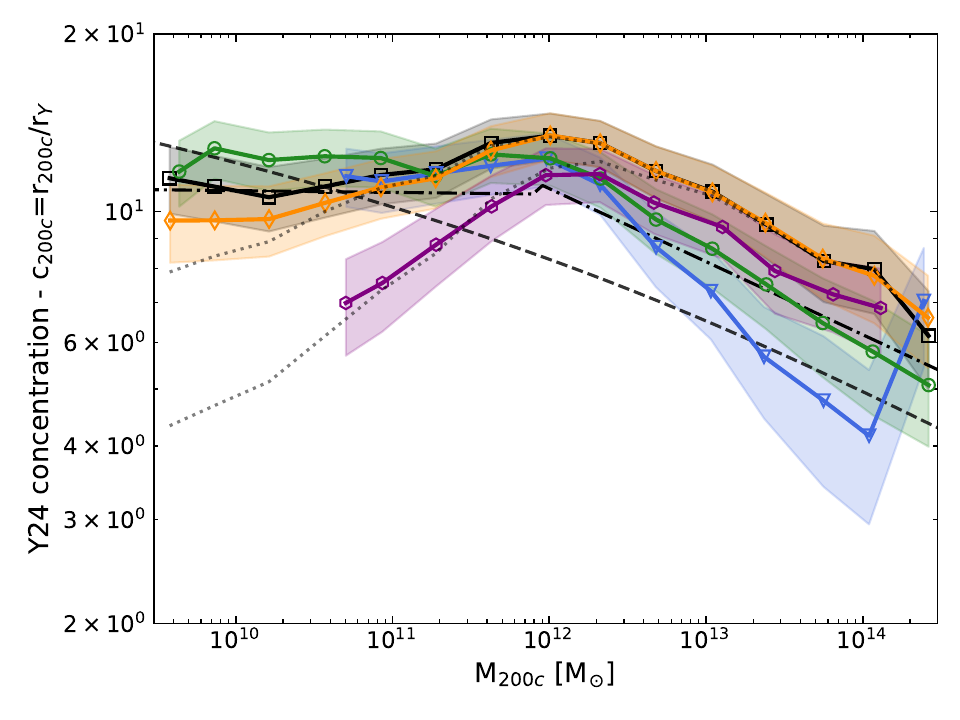}
    \caption{Same as Fig. \ref{fig:ein_mean0} but for the dark matter profiles in the FP runs. We fit all profiles with the Einasto and Y24 models, calculate the running means. For Y24, instead of plotting r$_{c}$/r$_{200c}$, here we show the ratio between the FP best-fit value and the DMO one, to highlight the changes brought in by the presence of baryons. The bottom panels show the concentration-mass relation in the FP runs.} \label{fig:bar_effect}
\end{figure*}

\subsection{Density profiles and concentrations} 

We now analyse the dark matter density profiles in the FP AIDA–TNG simulations, repeating the analysis described in Sect. \ref{sec:dmo} for the DMO runs. As discussed in the previous section, the inclusion of baryonic physics significantly modifies the inner halo structure, leading to systematically steeper density profiles across all dark matter scenarios. As a consequence, the best-fit profile parameters shift and the need for explicitly cored analytic models is drastically reduced. Figure \ref{fig:bar_effect} shows the distribution of Einasto and Y24 best-fitting parameters for the FP runs, which can be directly compared to their DMO counterparts shown in Fig. \ref{fig:ein_mean0}. Across all models, the dark matter responds to the presence of baryons in a qualitatively similar manner, though the strength of the effect differs. At low masses, baryons constitute only a small fraction of the total mass and therefore induce minimal modifications to the density profiles. The strongest baryonic impact occurs around $M_{200c}\simeq10^{12}M_{\odot}$, where the central stellar component dominates the gravitational potential and steepens the inner dark matter slope. This behaviour was already visible in the vSIDM panel of Fig. \ref{fig:prof}, where haloes below and above $10^{12}M_{\odot}$ display markedly different inner structures. At higher masses, the decreasing stellar mass fraction weakens the baryonic back-reaction, and the density profiles across models converge again.

\begin{figure}
    \centering
    \includegraphics[width=\linewidth]{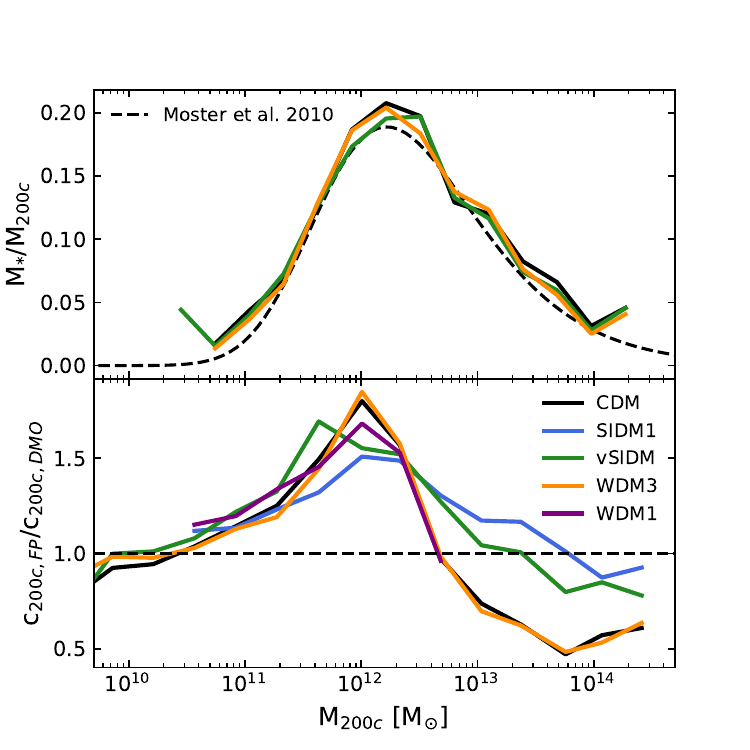}
    \caption{We show the correspondence between the trend in the stellar mass fraction as a function of halo mass measured in our runs (top) and the ratio between the FP and DMO Einasto concentration-mass relations at $z=0$ (bottom).}
    \label{fig:mstar_conc}
\end{figure}

The steepening of the inner profile suppresses the development of SIDM-induced constant-density cores in the FP runs. As a result, the Y24 core radius becomes very small at all masses, and the Y24 functional form loses its advantage over Einasto: as confirmed by the $\chi^2$ distributions reported in Appendix \ref{sec:app1}, the Einasto model becomes the preferred description even for SIDM1 and vSIDM haloes, except at the low-mass end $M_{200c}\leq10^{11.5}$$M_{\odot}$. As discussed in the previous section, the FP SIDM1 haloes in the mass range $M_{200} = [10^{12},\, 10^{13}]{M_\odot}$ show a particularly wide diversity of inner slopes, including systems with very steep cusps ($\gamma<-1.5$). This is evident by the peak in scale density and the corresponding dip in scale radius for both profiles. The same scale also shows an increase in concentration, as we see in the bottom panels of Fig. \ref{fig:bar_effect}. Here we show the concentration–mass relation for the FP CDM run and compare it with previous TNG-based studies. We reproduce the results of previous works \citet{lovell18,anbajagane22,sorini24}, who found that TNG baryonic physics produces a sharp enhancement in concentration around $M_{200c}\simeq 10^{12}M_{\odot}$. Remarkably, the same feature appears in all dark matter models, demonstrating that the baryonic impact on halo structure is largely independent of the underlying dark matter microphysics at this mass scale. This mass range also corresponds to the maximum of the stellar-to-halo mass relation of the AIDA runs \citep{despali25a}, in agreement with earlier findings that variations in concentration correlate tightly with the stellar mass fraction \citep{dicintio14b,dicintio14a}. In Fig. \ref{fig:mstar_conc}, we show this connection explicitly: the peak of the stellar mass fraction mirrors the peak in the ratio of FP-to-DMO mean concentrations, confirming a direct link between baryonic assembly efficiency and the strength of adiabatic contraction effects.

\section{Redshift evolution} \label{sec:redshift}

\begin{figure*}
    \centering
    \includegraphics[width=0.8\textwidth,valign=t]{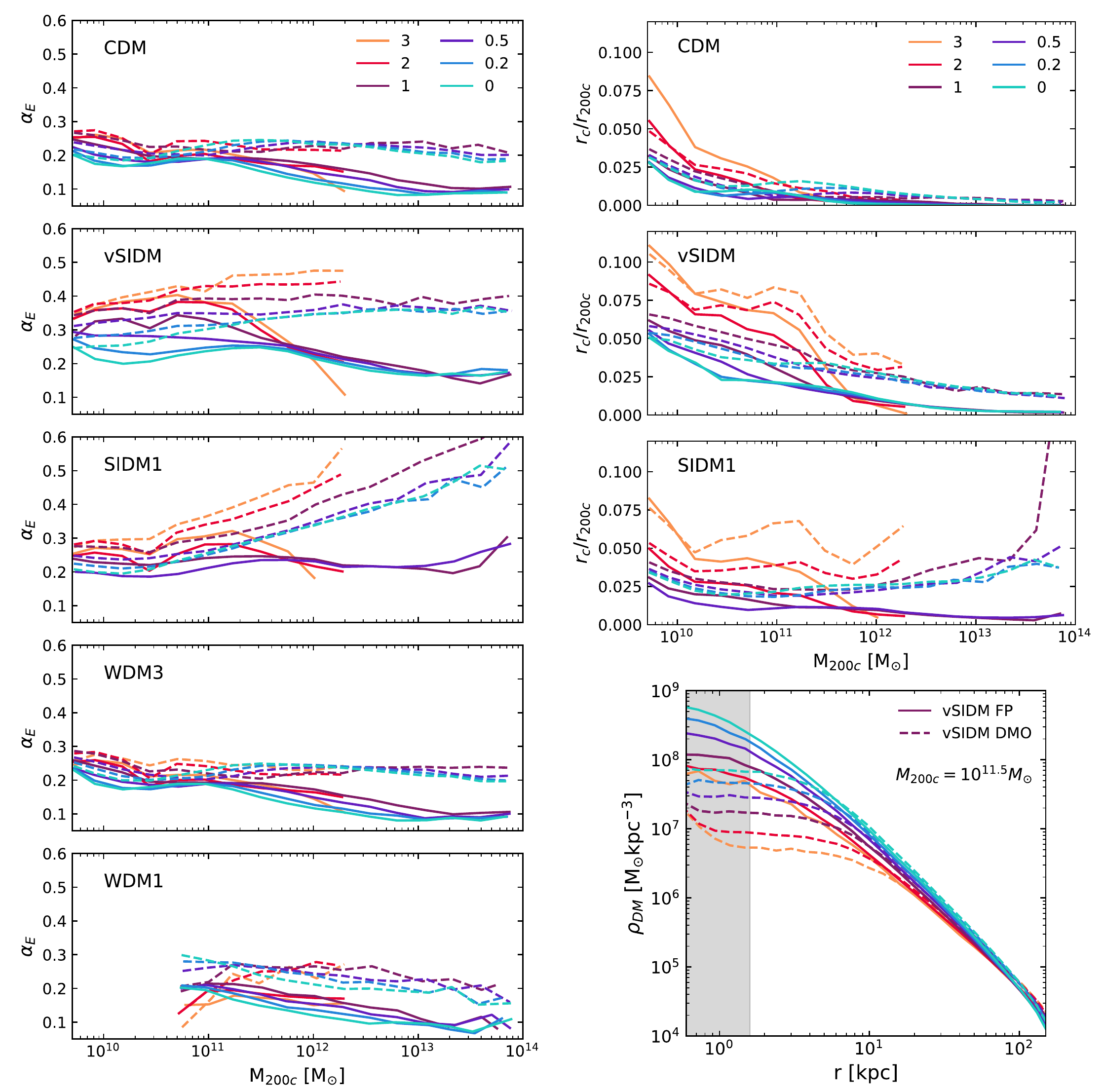}

    \caption{Redshift dependence of $(i)$ the Einasto parameter $\alpha_{E}$ (left) and $(ii)$ the ratio between the core size $r_{c}$ of the Y24 profile and the virial radius $r_{200c}$  (right). For the latter, we only show CDM, SIDM1 and vSIDM. Dashed (solid) lines represent the mean of DMO (FP) runs, as a function of halo mass. In the bottom-right panel, we show an example of redshift dependence of the dark matter profile: we plot the mean profile of haloes of mass $M_{200c}=10^{11.5}M_{\odot}$ for the vSIDM scenario, comparing the DMO (dashed ) and FP (solid) cases.}\label{fig:rc_z}
\end{figure*}

We expand the analysis described in the previous sections, considering five additional snapshots between $z=0.2$ and $z=3$, in order to study the redshift dependence of the dark matter profile properties and their model parameters that better describe the simulation data.  It is important to stress that here we look at the redshift dependence, or evolution, of the best-fit parameters, in a defined mass bin, and thus do not follow haloes along their merger histories, but show the properties of haloes at different redshifts.

At higher redshifts, we recover the same qualitative behaviour for the WDM and SIDM models as at $z=0$. We also reproduce the well-known redshift dependence of the concentration-mass relation found in previous works \citep{duffy08,maccio08,ludlow16,zhao09,giocoli12b}: the mean halo concentration decreases with redshift at fixed mass. However, it is interesting to discuss the redshift dependence of the characteristic physical scales that describe the density profiles, expressed by the best-fit parameters of the Einasto and Y24 models. In both cases, the scale radius and density evolve with redshift (the former decreasing and the latter increasing) at fixed mass, following the build-up of the profiles via accretion and the growth of haloes. The size of the scale radius (at fixed halo mass) shrinks as $\sim(1+z)$ - so that $r_{s}/(1+z)$ is roughly universal. Conversely, $\rho_{s}$ increases towards $z=0$ due to the build-up of haloes via merging. This determines the increase in halo concentration towards $z=0$ at all masses, given that $R_{200c}$ has a slower evolution at fixed mass \citep{diemer13}.  For WDM, we find that the turnover predicted by \citet{lovell14} and \citet{ludlow16} provides a good description of the concentration-mass relation in AIDA at all redshifts. This is consistent with the fact that the turnover is determined by the intrinsic properties of each WDM model (particle mass or free-streaming scale) and not by the halo evolution. In SIDM, the formation of the inner core is a product of particle interactions over time, coupled with the underlying growth of the halo that causes the increase of $r_{200c}$ at fixed mass. 

In Fig. \ref{fig:rc_z}, we show the redshift dependence of the third parameter of both fits: $\alpha_{E}$ and r$_{c}$. In the left panels, we see that the mean $\alpha_{E}$ as a function of halo mass shows almost no redshift dependence in WDM and CDM. Moreover, the addition of baryons (results presented using solid lines) reduces the value of $\alpha_{E}$ for masses $\geq10^{11}M_{\odot}$ compared to the DMO case (using dashed lines), consistently with the steepening of the inner profile due to adiabatic contraction. 
Self-interactions are instead a physical process that modifies the matter distribution over time. Thus, the redshift evolution of SIDM1 and vSIDM inner profiles is more visible: $\alpha_{E}$ increases with redshift, similarly to the core radius shown in the right panels as $r_{c}/r_{200c}$. The comoving size of $r_{c}$ decreases towards $z=0$, indicating that core formation already starts at high redshift and the core then evolves with the increasing density inside haloes towards the present time. The evolution of the core radius, at fixed mass, is slower than $r_{Y}$ and, as a consequence, the core occupies a larger fraction of the region within the scale radius at $z=0$. As an example of this behaviour, in the bottom-right panel of Fig. \ref{fig:rc_z}, we show the mean dark matter profile of haloes of mass $M_{200c}=10^{11.5}M_{\odot}$ at different redshifts in the vSIDM scenario, where we can appreciate the redshift evolution of the DMO and FP runs, presented using dashed and solid curves, respectively.

\section{Summary and conclusions} \label{sec:conc}

In this work, we investigate the dark matter density profile of haloes in the AIDA-TNG simulations \citep{despali25a}, considering systems with total halo mass between $5\times10^{9}M_{\odot}$ and $5\times10^{14}M_{\odot}$. This analysis is the first to address models for the density profiles in CDM, WDM, and SIDM models over such a wide range of halo masses, including the effects of full physical baryon processes. This is possible thanks to the setup of the AIDA-TNG sample that includes cosmological magneto-hydrodynamical simulations of the same boxes, considering multiple alternative dark matter models: warm dark matter (WDM) with particle masses of 1 and 3 keV (WDM1, WDM3), and self-interacting dark matter with either a constant (SIDM1) or velocity-dependent (vSIDM) scattering cross-section.
We measure the halo dark matter profiles in all runs, study the best-fit analytical models for each scenario, and examine the concentration-mass relation with and without the inclusion of baryons.  To ensure the robustness of the results, we only trust radial bins  $r\geq3\epsilon_{\rm DM}$, where $\epsilon_{\rm DM}$ is the Plummer-equivalent softening length that is used to quantify the simulation resolution. This corresponds to $r\geq4.4$ kpc for the 100/A runs and $r\geq1.7$ kpc for the 50/A runs.

In Sect. \ref{sec:dmo}, we first look at the DMO runs, in a way to isolate the impact of the dark matter physics on halo structures. 
In WDM models, the dark matter profiles are well described by the Einasto model \citep{einasto65}, with no particular evidence of central cores. The suppression of small-scale power leads to systematically lower concentrations at low masses consistent with the results of previous works \citep{lovell14,ludlow16}. On the other side, in the DMO SIDM models, with constant (SIDM1) and velocity-dependent (vSIDM) cross-section, the dark matter profiles form large and flat inner cores, which are best described by an analytical model which explicitly includes a core, such as the Y24 profile \citep{yang24}. The core size increases with halo mass but presents a different dependence on mass in the two models: vSIDM haloes form relatively larger cores in low-mass systems (due to higher cross-section at low velocities), while SIDM1 haloes produce larger cores at the high-mass end. It is important to underline that in both cases, the size of the core radius remains smaller than the scale radius of the profile, meaning that the core does not affect halo concentrations, which are essentially unchanged from CDM as shown in Fig. \ref{fig:ein_mean0}. This implies that a complete description of the structure of SIDM haloes requires combining the concentration-mass relation with a model for the core size (see Eq. \ref{eq:model}). 

In Sect. \ref{sec:hydro}, we investigate how these results change in the FP runs that include baryonic physics. We show that adiabatic contraction is stronger in self-interacting models for haloes of mass $M_{200c}\geq10^{11-11.5}M_{\odot}$, because the presence of the central galaxy prevents the formation of the dark matter core or strongly reduces it. As a consequence, the inner slope of the profile is shallower than CDM at the low-mass end, but then bends back to steeper values for galaxies and groups. Moreover, as shown in Fig. \ref{fig:prof_slope}, the interplay between self-interactions and galaxy formation produces a larger diversity of inner slopes compared to CDM, especially at the scale of Milky Way galaxies. Baryonic physics thus leaves strong imprints on the dark matter profiles, modifying the distribution of best-fit parameters. In particular, as SIDM cores disappear or become less flat, the FP profiles are best fit by the Einasto model rather than the Y24 cored profile.

Finally, in Sect. \ref{sec:redshift}, we extended our analysis to higher redshifts by repeating the profile fitting at five additional snapshots ($z=0.2,\,0.5,\,1,2\,,3$). The redshift evolution reinforces the picture emerging at  $z=0$: baryonic effects become progressively weaker at earlier times when galaxies are less massive, allowing SIDM cores to persist longer in lower-mass haloes. At later times, as galaxies grow and central potentials deepen, baryons increasingly dominate the inner profile, erasing cores and narrowing the structural differences between CDM, WDM, and SIDM at fixed mass.

Overall, the AIDA-TNG simulations highlight that the impact of alternative dark matter physics on the halo internal profile and structure cannot be interpreted without accounting for baryonic effects, their mass and redshift dependence.  
While DMO simulations reveal the characteristic signatures of WDM and SIDM, the FP runs show that galaxy formation can suppress, enhance, or even invert these trends, particularly near the peak of the stellar-to-halo mass relation. Understanding the combined imprint of baryons and dark matter microphysics will therefore be crucial for future observational probes of halo structure.

The diversity of inner slopes, the mass dependence of SIDM core suppression by baryons, and the characteristic scale at which concentration is maximally enhanced by galaxy formation all offer measurable signatures that can be targeted through joint weak and strong lensing analyses, stellar kinematics, and resolved gas and X-ray profiles. Looking ahead, the structural differences uncovered in this work will soon enter a regime that will become observationally testable. Forthcoming multiwavelength datasets -- from deep optical and near-infrared imaging with Euclid \citep{Euclid_report,euclidoverview,scaramella22} and Rubin \citep{ivezic08,ivezic09,lsst}, to high-resolution X-ray and SZ measurements from Athena \citep{athena12,athena20,athena25} and the new generation of millimetre observatories -- will provide unprecedented constraints on the inner density structure of galaxies and groups. Our results, therefore, provide a physically motivated framework for interpreting these observations in alternative dark matter models, emphasising the importance of combining probes across wavelengths, redshifts, and mass scales. As the quality and volume of data improve, the synergy between simulations such as the AIDA-TNG and the upcoming observational landscape will make it possible to robustly trace and label the imprint of warm or self-interacting dark matter in the structure of galaxies and their haloes.

\section*{Data Availability}

The original IllustrisTNG simulations are publicly available and accessible at \url{www.tng-project.org/data} \citep{nelson19}, where AIDA-TNG will also be made public in the future. Data directly related to this publication is available on request from the corresponding author. Additional information about AIDA can be found online at \url{https://gdespali.github.io/AIDA/}.

\begin{acknowledgements}
GD acknowledges the funding by the European Union - NextGenerationEU, in the framework of the HPC project – “National Centre for HPC, Big Data and Quantum Computing” (PNRR - M4C2 - I1.4 - CN00000013 – CUP J33C22001170001). We acknowledge the EuroHPC Joint Undertaking for awarding this project access to the EuroHPC supercomputer LUMI, hosted by CSC (Finland) and the LUMI consortium through a EuroHPC Extreme Scale Access call. We acknowledge ISCRA and ICSC for awarding this project access to the LEONARDO supercomputer, owned by the EuroHPC Joint Undertaking, hosted by CINECA (Italy). LM acknowledges the financial contribution from the 
PRIN-MUR 2022 20227RNLY3 grant “The concordance cosmological model: stress-tests with galaxy clusters” supported by Next Generation EU and from the grant ASI n. 2024-10-HH.0 “Attività scientifiche per la missione Euclid – fase E”. AP acknowledges funding by the European Union(ERC, COSMIC-KEY, 101087822, PI: Pillepich)

\\This research made use of the public Python packages matplotlib \citep{matplotlib}, NumPy \citep{numpy} and \textsc{COLOSSUS} \citep{diemer18}.

\end{acknowledgements}

\bibliographystyle{aa}
\bibliography{aanda.bbl}

\begin{appendix}
\onecolumn

\section{Fitting WDM and SIDM density profiles} \label{sec:app1}

\begin{figure*}[!h]
    \centering
    \includegraphics[width=\textwidth]{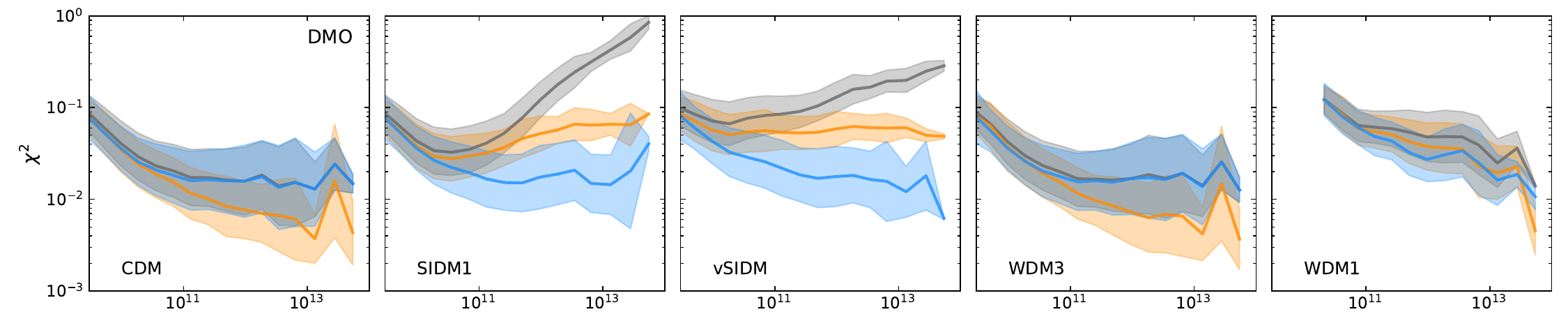}
    \includegraphics[width=\textwidth]{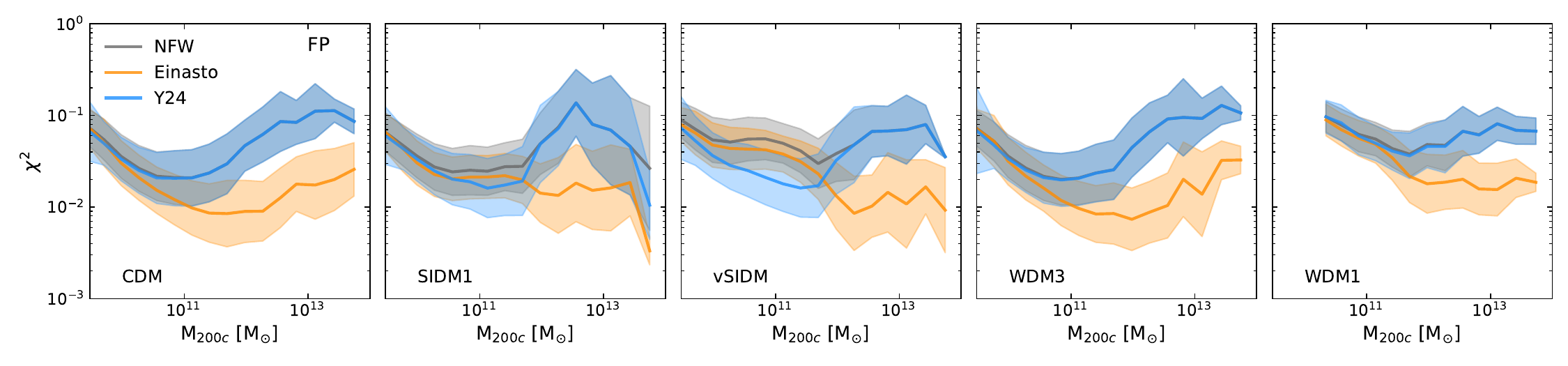}
    \caption{Mean goodness of fit (quantified by the $\chi^2$) of the NFW (grey), Einasto (orange), and Y24 (blue) models in each considered dark matter scenario; we show the DMO and FP runs in the top and bottom row, respectively.  In each panel, the solid lines show the mean value as a function of mass for the 50/A resolution, except for WDM1, where 50/B was considered, while the shaded band indicates the standard deviation.} \label{fig:chi2}
\end{figure*}

We now provide a quantitative estimate of the goodness of fit of the three considered analytical models, given by the distribution of $\chi^{2}$ values calculated as 
\begin{equation}
\chi^{2}=\frac{1}{N-n_{dof}}\sum_{1}^{N}\frac{(\log\rho_{model}-\log\rho_{data})^{2}}{\log\rho_{data}} .
\end{equation}
In Fig.\ref{fig:chi2} we show the distribution of $\chi^2$ values for the NFW, Einasto, and Y24 models in each dark matter scenario. The $\chi^2$ is calculated only for radii $3\epsilon_{DM}\leq r\leq r_{200c}$, to avoid fitting for the artificial inner core due to resolution or the outskirts of the halo where the density can be affected by the environment or the halo identification algorithm. The top row panels show the goodness of fit in the DMO runs: for CDM and WDM, the Y24 and Einasto models have the same performance in the 50/A high-resolution runs, consistently with the definition of the Y24 profile that reverts to NFW in the absence of a core; at the same time, the Einasto model is a better fit, especially at the high-mass end. Instead, in the two SIDM models, the halo profiles are more accurately described by the Y24 profile, followed by the NFW profile. 

\section{Convergence tests for SIDM}  \label{sec:app2}
It is well-known that the finite spatial resolution of simulations introduces important limitations on the reliability of the inner part of measured density profiles. Many previous works have explored these limitations and performed convergence tests in CDM, especially in DMO simulations \citep{springel08b}. It is thus common practice to use multiples of the gravitational softening $\epsilon_{DM}$ to determine the minimum scale where the dark matter profile is reliably measured, usually in the range $r\geq2.3-3\epsilon_{DM}$. At smaller radii, an artificial core tends to be created by numerical effects. In FP simulations, the presence of stars at the centre of haloes influences the central density and its slope, making it deviate from the standard NFW or Einasto profile (see Fig. \ref{fig:prof}). This effect is not identical in all mass bins, but strongly depends on the stellar mass fraction that sits at the centre, and it is thus particularly important in the mass range $M_{200c}\sim10^{12}-10^{13.5}M_{\odot}$. It will also depend on resolution: as discussed in \citet{pillepich18b}, stellar masses and star formation rates typically increase with better resolution in the TNG model. This is because the model parameters remain unchanged in all runs, without intentional adjustments for different resolutions. However, \citet{pillepich18b} also demonstrated that the ancillary runs (such as our 100/B and 50/B boxes) can be used to model these effects and introduce useful rescaling parameters.

\begin{figure*}
    \centering
    \includegraphics[width=0.9\textwidth]{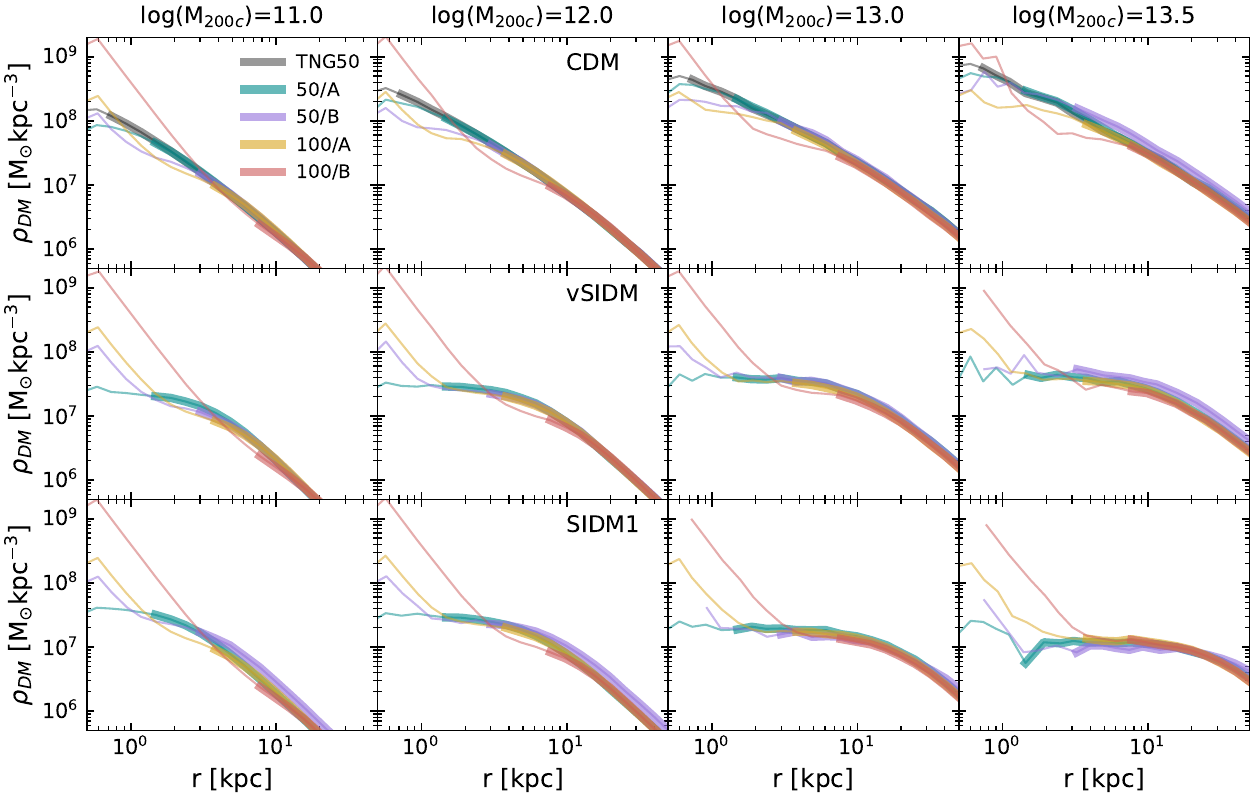}
    \includegraphics[width=0.9\textwidth]{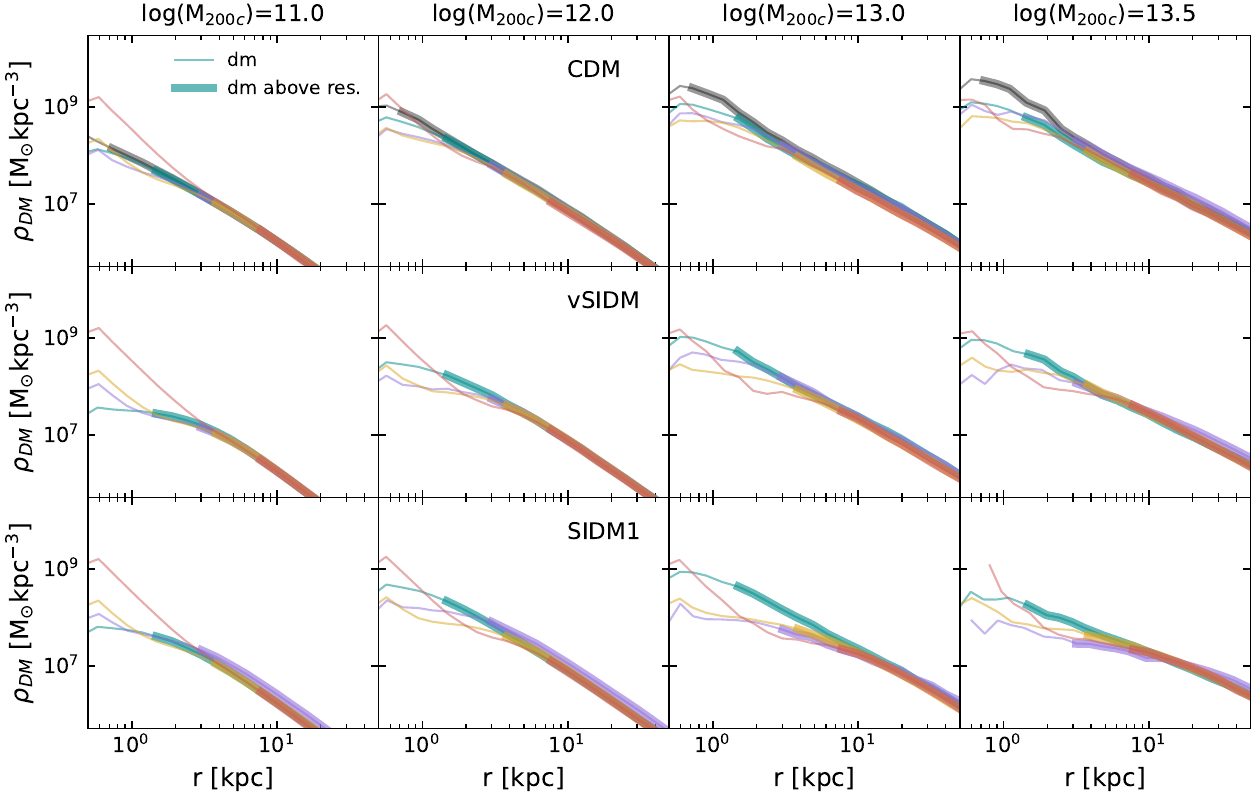}
    \caption{Convergence tests for the dark matter profile in the 50/A boxes in the DMO (top) and FP (bottom) runs. We consider CDM, SIDM1 and vSIDM and we do not show results for the WDM models, as these behave similarly to CDM. For comparison, we also show results from the IllustrisTNG 50 Mpc run, since its resolution is higher than our best case. We plot the mean dark matter profile for four representative bins in halo mass of width 0.2 dex around the reported value. In each panel, the thick solid lines show the density profiles up to the resolution limit of each simulation, while the thinner lines help visualise where convergence breaks at lower radii. } \label{fig:res1}
\end{figure*}

\begin{figure*}
    \centering
    \includegraphics[width=0.9\textwidth]{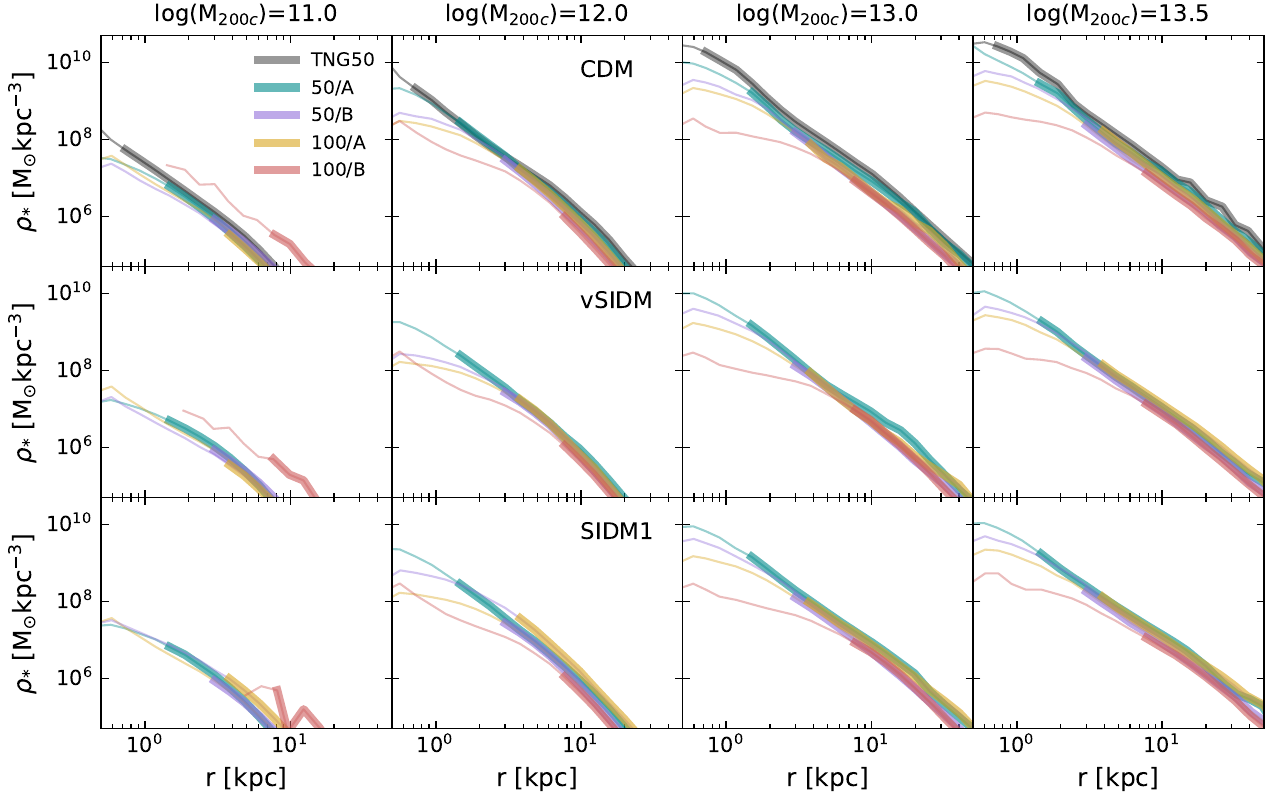}
    \caption{Same as for Fig. \ref{fig:res1} but for the stellar profiles. As reported by \citet{pillepich18b}, the stellar mass increases with resolution, leading to a different level of adiabatic contraction. } \label{fig:res2}
\end{figure*}

We recover these results in our CDM runs and find similar behaviours also in WDM, given that the particle interactions are not fundamentally different and the density profiles of haloes with a relevant stellar component are not affected in the considered models (see Fig. \ref{fig:ein_mean0}). However, self-interacting models introduce non-gravitational forces between particles, which lead to the formation of an inner core. It is thus worth studying the numerical convergence in these models and the interplay between self-interactions and baryonic effects. Given that the latter causes an increase in the central dark matter density due to adiabatic contraction, we need to test if the SIDM core reacts similarly at all resolutions. Fig.\ref{fig:res1} shows convergence tests on the dark matter profiles of haloes in four representative mass bins between $M_{200c}=10^{11}M_{\odot}$ and $M_{200c}=10^{13.5}M_{\odot}$ in the 50/A boxes. Given that the 51.7 Mpc boxes only include a few haloes with higher masses, we do not include them in this plot. In CDM, we reproduce the results of previous convergence tests in DMO runs and find good agreement between the runs for $r\geq3\times\epsilon_{DM}$. In vSIDM and SIDM1, we find that the DMO dark matter core is well reproduced at all resolutions, with a convergence level similar to or higher than CDM. As discussed in \citet{adhikari22}, the reason for this is that numerical effects in CDM simulations typically form cores, so when the physical model itself is one where a core is expected to form, these numerical effects are less significant. In fact, one of the drivers of non-convergence in the inner regions of CDM haloes is the spurious two-body gravitational scattering between particles, which adds only a small perturbation to SIDM, which has a high physical cross section for particle-particle scattering.  In the FP case (bottom), we see the reaction of dark matter to the presence of baryons, which is especially relevant in the two central columns, where the stellar mass fraction at the centre is the highest. 
In CDM and vSIDM, the dark matter FP profiles still converge well at all masses above the resolution limit, with some departures at the high-mass end. This is not the case in SIDM1, where we observe departures up to larger radii, peaking at $M_{200c}=10^{12-13}M_{\odot}$. Figure \ref{fig:res2} shows instead the density profiles of stars in CDM, SIDM1 and vSIDM for the same of resolution levels of Fig. \ref{fig:res1}, showing that the overall slope and shape of the profiles are similar in all dark matter models, but the stellar mass increases with resolution, as reported in \citet{pillepich18b}. In turn, this increase may have a different effect on self-interacting models compared to CDM.

\end{appendix}
\end{document}